\documentclass[journal]{IEEEtran}
\usepackage{}
\usepackage{amsfonts}
\usepackage{savesym}
\usepackage{amsmath}
\savesymbol{iint}
\restoresymbol{TXF}{iint}
\usepackage{amssymb}
\usepackage{cases}
\usepackage{booktabs}
\usepackage{latexsym}
\usepackage{psfrag}
\usepackage{graphicx,subfigure}
\usepackage{hyperref}
\usepackage{color}
\usepackage{bm}
\usepackage{amsfonts,mathrsfs,bbm,amsmath,stmaryrd,amssymb,pifont}%
\usepackage{amscd,graphicx,array,dsfont,texdraw,tikz}%
\usetikzlibrary{arrows,decorations.markings,shapes,shadows,fadings}
\usepackage{bbm,mathrsfs}%
\usepackage{array}
\usepackage{epstopdf}
\usepackage[linesnumbered,ruled,vlined]{algorithm2e}
\allowdisplaybreaks[4]
\definecolor{blue}{rgb}{0,0,0}
\newtheorem{theorem}{Theorem}
\newtheorem{lemma}{Lemma}

\newtheorem{remark}{Remark}
\newtheorem{assumption}{Assumption}

\newtheorem{corollary}{Corollary}

\begin{document}

\title{Distributed Scalable Coupled Policy Algorithm for Networked Multi-agent Reinforcement Learning}

\author{Pengcheng Dai, Dongming Wang, Wenwu Yu~\IEEEmembership{Senior Member,~IEEE}, and Wei Ren~\IEEEmembership{Fellow,~IEEE,}
\thanks{
This work was supported in part by the National Key Research and Development Program of China under Grant 2022ZD0120002, in part by the Jiangsu Provincial Key Laboratory of Networked Collective Intelligence under Grant BM2017002, and in part by the National Natural Science Foundation of China under Grant Nos. 62233004.
\emph{(Corresponding author: Wenwu Yu.)}}
\thanks{Pengcheng Dai is with the Engineering Systems and Design Pillar, Singapore University of Technology and Design, Singapore 487372 (e-mail: Jldaipc@163.com).}
\thanks{Dongming Wang and Wei Ren are with the Department of Electrical and Computer Engineering, University of California, Riverside, CA 92521, USA. (e-mail: wdong025@ucr.edu; ren@ece.ucr.edu).}
\thanks{Wenwu Yu is with the Frontiers Science Center for Mobile Information Communication and Security, School of Mathematics, Southeast University, Nanjing 210096, China, and also with Purple Mountain Laboratories, Nanjing 211102, China (e-mail: wwyu@seu.edu.cn).}
}

\markboth{Manuscript For
 Review}
{Manuscript For Review}

\maketitle

\begin{abstract}
This paper studies networked multi-agent reinforcement learning (NMARL) with interdependent rewards and coupled policies. In this setting, each agent's reward depends on its own state-action pair as well as those of its direct neighbors, and each agent's policy is parameterized by its local parameters together with those of its $\kappa_{p}$-hop neighbors, with $\kappa_{p}\geq 1$ denoting the coupled radius.
The objective of the agents is to collaboratively optimize their policies to maximize the discounted average cumulative reward.
To address the challenge of interdependent policies in collaborative optimization, we introduce a novel concept termed the ``neighbors' averaged $Q$-function'' and derive a new expression for the coupled-policy gradient.
Based on these theoretical foundations, we develop a distributed scalable coupled policy (DSCP) algorithm, where each agent relies only on the state-action pairs of its $\kappa_{p}$-hop neighbors and the rewards of its $(\kappa_{p}+1)$-hop neighbors.
{\color{blue}Specifically}, in the DSCP algorithm, we employ a geometric 2-horizon sampling method that does not require storing a full $Q$-table to {\color{blue}obtain an unbiased estimate of the surrogate coupled-policy gradient under the executed policy.}
Moreover, each agent interacts exclusively with its direct neighbors to obtain accurate policy parameters, while maintaining local estimates of other agents' parameters to execute its local policy and collect samples for optimization.
These estimates and policy parameters are updated {\color{blue}in a distributed manner} through local information exchange among direct neighbors, enabling coordinated policy updates across the network.
{\color{blue}We prove that the gradient norm of the objective function along the joint policy sequence produced by the proposed algorithm converges to zero almost surely, i.e., the algorithm is asymptotically first-order stationary.}
Finally, the effectiveness of {\color{blue}the} DSCP algorithm is demonstrated through simulations in a robot path planning environment, {\color{blue}showing clear improvement over the compared algorithms in the tested setting.}
\end{abstract}


\IEEEpeerreviewmaketitle

\section{Introduction}
With the advancement of artificial intelligence technology, reinforcement learning (RL)~\cite{Sutton1998} has garnered increasing attention from both academia and industry.
Currently, multi-agent reinforcement learning (MARL) has demonstrated outstanding performance {\color{blue}in} many complex scenarios, such as path planning~\cite{DaiTAC2025,ZhouUAI2023} and
control and optimization {\color{blue}systems}~\cite{Sha2022,Duan2016,Levine2016}.
\par
Network communication can substantially enhance multi-agent decision-making, which has led to the formulation of MARL over networks (i.e., networked MARL, NMARL).
In the NMARL problem, agents leverage the network to exchange their local information (e.g., {\color{blue}state-action pairs, rewards, and learned parameters}) with their neighbors to achieve cooperative optimization.
\subsection{Related literature}
Early studies primarily focused on fully connected interaction structures or assumed that agents had access to globally aggregated information.
Notable examples include centralized training with decentralized execution frameworks, such as value-decomposition approaches (e.g., VDN~\cite{VDN}, QMIX~\cite{QMIX}, QPLEX~\cite{QPLEX}), and cooperative actor-critic methods (e.g., COMA~\cite{COMA}).
While these algorithms provide strong coordination performance, they rely on fully connected networks and centralized computation, which become increasingly impractical in large-scale systems.
\par
{\color{blue}To address the scalability issue} in large-scale systems, mean-field MARL approaches~\cite{YangICML2018,LuoTCYB2021} have recently been developed within the NMARL framework, leveraging communication networks.
In these methods, each agent approximates the influence of others via the average behavior of its neighbors, thereby reducing dependence on global action information.
Nevertheless, these approaches still require all agents to access a shared global state, which restricts their applicability in large-scale or communication-constrained environments.
\par
To further reduce the algorithm's dependence on the global state-action {\color{blue}information} in the NMARL problem, a truncated $Q$-function technique was proposed in~\cite{QuCLDC2020,QuNIPS2020}, showing that each agent's $Q$-function can be approximated using only the state-action pairs within its $\kappa$-hop neighborhood.
Leveraging this property, several scalable algorithms were developed in~\cite{QuCLDC2020,QuNIPS2020}, where each agent relies solely on the state-action pairs of its $\kappa$-hop neighbors.
Even so, the $Q$-functions are represented {\color{blue}in tabular form}, which requires substantial storage as the communication range increases.

\subsection{Motivation}
While the aforementioned algorithms have achieved significant progress in addressing the NMARL problem, they still face two major limitations.
The first limitation arises from scalable NMARL methods that employ truncated $Q$-functions~\cite{QuCLDC2020,QuNIPS2020}.
In these approaches, the $Q$-functions are represented in tabular form, which introduces substantial storage requirements as the truncated range increases.
Moreover, any approximation errors in the tabular $Q$-function will propagate to the policy gradient, potentially degrading the quality of the learned policy.
The second limitation is that most existing algorithms assume that agents act according to independent policies.
This assumption, however, is often unrealistic in practical scenarios, where agents' decisions are inherently interdependent.
In practical applications such as traffic signal control~\cite{ChuTITS2020,DaiTII2024}, algorithms based on coupled policies typically exhibit significantly stronger learning capability than those relying on independent policies.
The main reason is that the traffic dynamics at one intersection are tightly linked to the policies of neighboring intersections, and ignoring such coupling prevents intersections from collaboratively optimizing traffic flows.
{\color{blue}Although coupled-policy algorithms have demonstrated promising empirical performance, their theoretical foundations remain largely unexplored.}
\par
Motivated by these challenges, this paper addresses the following question:~\emph{How {\color{blue}can we} design a distributed and scalable algorithm for NMARL with coupled policies that avoids excessive computational overhead while ensuring convergence guarantees?}
\par
\subsection{Contribution}
To address the aforementioned challenges, this paper makes the following key contributions.
\par
(i) \textbf{NMARL problem formulation with coupled policies}: We formulate an NMARL framework with coupled policies, where each agent's local policy is influenced not only by its own parameter but also by those of its $\kappa_{p}$-hop neighbors.
To eliminate the need for global state-action information, a novel concept termed the ``neighbors' averaged $Q$-function'' is introduced, based on which a new form of the coupled-policy gradient is derived (see Theorem~\ref{thelemmainNNMARLITP}).
Leveraging this gradient, we develop a distributed scalable coupled policy (DSCP) algorithm, where each agent relies only on the state-action pairs of its $\kappa_{p}$-hop neighbors and the rewards from its $(\kappa_{p}+1)$-hop neighbors.
\par
(ii) \textbf{Distributed algorithm with efficient sampling}: In {\color{blue}the} DSCP algorithm, we employ a geometric 2-horizon sampling method that does not require storing a full $Q$-table {\color{blue}to estimate the surrogate direction under the executed policy.}
Moreover, in the DSCP algorithm, each agent interacts only with its direct neighbors {\color{blue}to obtain their true} policy parameter and maintains its estimates of the policy parameters of other agents to execute its local policy and collect samples for optimization.
These estimates and the policy parameters are then updated in a {\color{blue}distributed manner}, enabling {\color{blue}coordinated} policy updates across the network.
\par
(iii) \textbf{Convergence guarantees}: We establish that each agent optimizes its local policy parameters using
{\color{blue}an unbiased estimate of the surrogate direction under the executed policy}
(see Lemma~\ref{thelemmaofapproximatedpolicygradient}), and its local estimates of other agents' policy parameters converge to their true values (see Theorem~\ref{thelemmaofpolicyparameterconvergence}).
Furthermore, {\color{blue}the gradient norm of the objective function along the joint policy sequence generated by the proposed DSCP algorithm converges to zero both almost surely and in expectation} (see Theorem~\ref{thetheoremconvergenceofpolicygradient}).
\par
The rest of this paper is organized as follows.
The notations and preliminary concepts related to network are presented in Section~\ref{SectionIIpreliminaries}.
The NMARL problem with coupled policies is described in Section~\ref{SectionIII}.
Section~\ref{thesectionalgorithmdesign} presents the design of the DSCP algorithm.
The convergence analysis of the proposed DSCP algorithm is provided in Section~\ref{SectionIV}.
Simulation results demonstrating the effectiveness of the DSCP algorithm are presented in Section~\ref{SectionSimulations}.
Finally, the conclusion and further work are discussed in
Section~\ref{SectionVConclusions}.

\section{Preliminaries}\label{SectionIIpreliminaries}
\subsection{Notations}\label{notationsinthispaper}
$\mathbb{R}$ is the set of reals.
$\mathbb{R}^{N}$ and $\mathbb{R}^{N\times M}$ represent
the $N$-dimensional vector set and the $N\times M$-dimensional matrix set, respectively.
For vectors $x, y\in\mathbb{R}^{N}$, $x^{\top}$ denotes the transpose of $x$, and $x \otimes y$ denotes the Kronecker product of vectors $x$ and $y$.
The operator $\|\cdot\|$ applied to vectors denotes the standard
$\mathcal{L}_{2}$-norm, while applied to matrices denotes the induced $\mathcal{L}_{2}$-norm.
The symbol $\mathbf{1}_{N}=(1,\cdots, 1)^{\top}$ denotes the $N$-dimensional column vector consisting of all ones, whereas $\mathds{1}_{\{\cdot\}}$ represents the indicator function.

\subsection{Network}\label{theintroduvtionofnetwork}
Let $\mathcal{G}(\mathcal{N}, \mathcal{E})$ denote {\color{blue}an undirected} network over $N$ agents, where $\mathcal{N} = \{1,\cdots, N\}$ is the set of agents and $\mathcal{E}$ is the set of edges among agents.
{\color{blue}An edge $(i,j) \in \mathcal{E}$ means that agents $i$ and $j$ can exchange information with each other. A sequence of agents $(i_1, i_2, \cdots, i_k)$ is called a path between agents $i_1$ and $i_k$ if $(i_l, i_{l+1}) \in \mathcal{E}$ for all $1 \leq l \leq k-1$. The network $\mathcal{G}(\mathcal{N}, \mathcal{E})$ is said to be connected if there exists a path between any two agents $i, j \in \mathcal{N}$. Denote $\mathcal{N}_i = \{j|(i,j) \in \mathcal{E}\} \cup \{i\}$ as the direct neighborhood of agent $i$, which includes agent $i$ itself, and $|\mathcal{N}_i|$ as its cardinality.}
For integer $\kappa_p \geq 1$, we let $\mathcal{N}_i^{\kappa_p}$ denote the $\kappa_p$-hop neighborhood of agent $i$, i.e., the agents whose graph distance to agent $i$ is less than or equal to $\kappa_p$, including $i$ itself.
Moreover, we define $\mathcal{N}_{-i}^{\kappa_p} = \mathcal{N} \setminus \mathcal{N}_i^{\kappa_p}$, and for any $i' \in \mathcal{N}$, denote $\mathcal{N}_{i,-i'}^{\kappa_p} = \mathcal{N}_i^{\kappa_p} \setminus \{i'\}$.
{\color{blue}Denote $W = [w_{ij}]_{N \times N}$ as the weight matrix of $\mathcal{G}(\mathcal{N}, \mathcal{E})$, which satisfies $w_{ij} = 1/|\mathcal{N}_j|$ if $i \in \mathcal{N}_j$, and $w_{ij} = 0$ otherwise.
}

\section{The NMARL problem with coupled policies}\label{SectionIII}
In this section, we present the formulation of the NMARL problem with coupled policies and introduce several preliminary results that will be used in the subsequent analysis.
\subsection{The NMARL problem setup}\label{ModelofNMARLproblem}
The model of the NMARL problem with coupled policies is described {\color{blue}by the tuple} $\big(\mathcal{G}(\mathcal{N},\mathcal{E}),\{\mathcal{S}_{i}\}_{i\in\mathcal{N}},\{\mathcal{A}_{i}\}_{i\in\mathcal{N}},$
$\{\mathcal{P}_{i}\}_{i\in\mathcal{N}},\{r_{i}\}_{i\in\mathcal{N}},\bm{\rho},\gamma\big)$,
where $\mathcal{G}(\mathcal{N},\mathcal{E})$ is the underlying network among agents.
$\mathcal{S}_{i}$ and $\mathcal{A}_{i}$ represent the local state space and the local action space of agent $i$, respectively.
Let $s_{i} \in \mathcal{S}_{i}$ denote the local state and $a_{i} \in \mathcal{A}_{i}$ denote the local action of agent $i \in \mathcal{N}$.
The global state $\bm{s}=(s_{1},\cdots,s_{N})$ and the global action $\bm{a}=(a_{1},\cdots,a_{N})$ are defined accordingly.
Let $\bm{s}_{t}=(s_{1,t},\cdots,s_{N,t})$ denote the global state at time $t$, and let $\bm{a}_{t}=(a_{1,t},\cdots,a_{N,t})$ represent the global action at time $t$.
Denote $s_{\mathcal{N}^{\kappa_{p}}_{i}}$ and $a_{\mathcal{N}^{\kappa_{p}}_{i}}$ as the states and actions of agent $i$'s $\kappa_{p}$-hop neighbors, respectively.
Likewise, $s_{\mathcal{N}^{\kappa_{p}}_{i},t}$ and $a_{\mathcal{N}^{\kappa_{p}}_{i},t}$ represent the states and actions of agent $i$'s $\kappa_{p}$-hop neighbors at time $t$, respectively.
$\mathcal{P}_{i}(s'_{i}|s_{i},a_{i})$ is the state transition probability function of agent $i$.
Define $\bm{\mathcal{P}}(\bm{s}'|\bm{s},\bm{a})=\prod_{i=1}^{N}\mathcal{P}_{i}(s'_{i}|s_{i},a_{i})$
as the global state transition probability function of agents.
$r_{i}(s_{\mathcal{N}_{i}},a_{\mathcal{N}_{i}})$ is the reward function of agent $i$, which depends on the state-action pairs of its direct neighbors.
For convenience, we define $r_{i,t}=r_{i}(s_{\mathcal{N}_{i},t},a_{\mathcal{N}_{i},t})$ as the instantaneous reward of agent $i$ at time $t$.
$\bm{\rho}$ is the distribution of the initial state $\bm{s}_{0}$ and $\gamma\in(0,1)$ is the discount factor.
\begin{remark}
The NMARL formulation described above follows a standard framework that has been widely studied in the literature~\cite{LiuarXiv2023}.
A representative motivating example is the distributed power control problem in wireless communication networks.
In this setting, each communication link
$i$ is modeled as an autonomous agent whose local state is typically given by its current transmission power level $0 \leq p_i \leq p_{\mathrm{max}}$, where $p_{\mathrm{max}}$ denotes the maximum allowable power.
The action $a_i\in\{0,-1,1\}$ indicates whether the agent maintains, decreases, or increases its transmission power by one unit.
The local reward of agent $i$ is then defined as
\begin{align}
r_{i}(s_{\mathcal{N}_{i}},a_{\mathcal{N}_{i}})
=\log\!\big(1+\frac{p_{i}g_{i,i}}{\sum_{j\in\mathcal{N}_{i}\setminus \{i\}} p_{j}g_{i,j}+\sigma_{i}}\big)-u_{i}p_{i},
\end{align}
where $g_{i,j}$ denotes the channel gain from agent $j$ to agent $i$, $\sigma_i$ represents the noise power at agent $i$, and $u_i$ is a weighting parameter that balances the trade-off between throughput and power consumption.
This model explicitly captures the interference coupling among neighboring communication links and illustrates the practical relevance of the proposed NMARL framework in wireless communication systems.
\end{remark}
\par
In light of the interplay among the decisions of agents, we formally define the local policy of agent $i$ as $\pi_{i}(\cdot|s_{i},\theta_{i},\theta_{\mathcal{N}^{\kappa_{p}}_{i,-i}})$, which depends not only on its local policy parameter but also incorporates the policy parameters of its $\kappa_{p}$-hop neighboring agents, where $\kappa_{p}\geq1$ is {\color{blue}a} constant.
Denote $\bm{\pi}_{\bm{\theta}}(\cdot|\bm{s})\triangleq\bm{\pi}(\cdot|\bm{s},\bm{\theta})=\prod_{i=1}^{N}\pi_{i}(\cdot|s_{i},\theta_{i},\theta_{\mathcal{N}^{\kappa_{p}}_{i,-i}})$ as the joint policy of all agents, where $\bm{\theta}=(\theta^{\top}_{1},\cdots,\theta^{\top}_{N})^{\top}$ is the joint policy parameter of all agents.
\par
In the NMARL problem, define the discounted average cumulative reward of agents under joint policy $\bm{\pi}_{\bm{\theta}}$ as
\begin{align}\label{theobjectivefunction}
J(\bm{\theta})=&\mathbb{E}_{\bm{s}\sim\bm{\rho}}\Big[\frac{1}{N}\!\sum_{t=0}^{\infty}\!\sum_{i=1}^{N}\gamma^{t}r_{i,t}|\bm{s}_{0}\!=\!\bm{s},\bm{a}_{t}\!\sim\!\bm{\pi_{\theta}}(\cdot|\bm{s}_{t})\Big].
\end{align}
The goal of all agents is to find the optimal joint policy parameter to maximize $J(\bm{\theta})$, i.e., $\max_{\bm{\theta}}J(\bm{\theta})$.
\par
For the purposes of the following development, we impose a structural assumption on the underlying network.
\begin{assumption}\label{theassumptionontheunderlyingnetwork}
In the NMARL problem, the underlying network $\mathcal{G}(\mathcal{N},\mathcal{E})$ is an undirected and connected network.
\end{assumption}
\par
The assumption of {\color{blue}an} undirected network is standard in NMARL~\cite{QuCLDC2020,QuNIPS2020,LiuarXiv2023} and is commonly adopted to characterize the communication relationships among agents.
\subsection{Preliminary results}
In the NMARL problem, for any joint policy $\bm{\pi_{\theta}}$,
define
the global $Q$-function $Q^{\bm{\pi_{\theta}}}(\bm{s},\bm{a})$ as
\begin{align}
Q^{\bm{\pi_{\theta}}}(\bm{s},\bm{a})=\mathbb{E}_{\bm{\pi_{\theta}}}\Big[\frac{1}{N}\sum_{t=0}^{\infty}\sum_{i=1}^{N}\gamma^{t}r_{i,t}|\bm{s}_{0}=\bm{s},\bm{a}_{0}=\bm{a}\Big],\label{thedefinitionofglobalQfunction}
\end{align}
and the local $Q$-function $Q^{\bm{\pi_{\theta}}}_{i}(s_{\mathcal{N}_{i}},a_{\mathcal{N}_{i}})$
of agent $i$ as
\begin{align}
Q^{\bm{\pi_{\theta}}}_{i}(s_{\mathcal{N}_{i}},a_{\mathcal{N}_{i}})=&\mathbb{E}_{\bm{\pi_{\theta}}}\Big[\sum_{t=0}^{\infty}\gamma^{t}r_{i,t}|s_{\mathcal{N}_{i},0}=s_{\mathcal{N}_{i}},a_{\mathcal{N}_{i},0}=a_{\mathcal{N}_{i}}\Big].\label{thelocalQfunctionofglobalpolicy}
\end{align}
Since the local reward function
$r_{i}(s_{\mathcal{N}_{i}}, a_{\mathcal{N}_{i}})$ of agent $i$
depends on the state-action pair $(s_{\mathcal{N}_{i}},a_{\mathcal{N}_{i}})$,
the domain of its corresponding local $Q$-function in (\ref{thelocalQfunctionofglobalpolicy}) naturally encompasses all possible combinations of $(s_{\mathcal{N}_{i}},a_{\mathcal{N}_{i}})$.
Moreover, owing to the independence of the local state transition probabilities
$\{\mathcal{P}_{i}(s'_{i}|s_{i}, a_{i})\}_{i\in\mathcal{N}}$,
a decomposition of the global $Q$-function can be established in the following lemma.
\begin{lemma}\label{thelemmaofdecomposition}
In the NMARL problem, for any joint policy $\bm{\pi_{\theta}}$, the global $Q$-function $Q^{\bm{\pi_{\theta}}}(\bm{s},\bm{a})$ can be decomposed as
\begin{align}
Q^{\bm{\pi_{\theta}}}(\bm{s},\bm{a})=\frac{1}{N}\sum_{i=1}^{N}Q^{\bm{\pi_{\theta}}}_{i}(s_{\mathcal{N}_{i}},a_{\mathcal{N}_{i}}).\label{theequationofDecompositionofvaluefunctions2}
\end{align}
\end{lemma}
\par
Lemma~\ref{thelemmaofdecomposition} achieves an exact decomposition of the global $Q$-function into the local $Q$-functions of individual agents.
This decomposition holds exactly without any truncation, {\color{blue}which follows from the factorized state transitions and policies in the NMARL problem, and thus differs from the truncated $Q$-function approach in~\cite{QuCLDC2020,QuNIPS2020}.}

\par
Building upon the decomposition result in Lemma~\ref{thelemmaofdecomposition}, we proceed to introduce the discounted state visitation distribution, a critical element in the derivation of the policy gradient theorem.
Define $d^{\bm{\pi_{\theta}}}_{\bm{\rho}}(\bm{s})$ as the discounted state visitation distribution, given by
\begin{align}\label{Thediscountedstatevisitationdistribution}
d^{\bm{\pi_{\theta}}}_{\bm{\rho}}(\bm{s})=(1-\gamma)\sum_{t=0}^{\infty}\gamma^{t}\mathrm{Pr}^{\bm{\pi_{\theta}}}(\bm{s}_{t}=\bm{s}|\bm{s}_{0}\sim\bm{\rho}),
\end{align}
where $\mathrm{Pr}^{\bm{\pi_{\theta}}}(\bm{s}_{t}=\bm{s}|\bm{s}_{0}\sim\bm{\rho})$ is the probability of occurrence of $\bm{s}_{t}=\bm{s}$ at time $t$ under the joint policy $\bm{\pi_{\theta}}$ and initial state distribution $\bm{\rho}$.
\par
Based on $d^{\bm{\pi_{\theta}}}_{\bm{\rho}}(\bm{s})$ in (\ref{Thediscountedstatevisitationdistribution}), the coupled-policy gradient $\nabla_{\theta_{i}}J(\bm{\theta})$ of agent $i$ is derived as follows.
\begin{theorem}\label{thelemmaofthepolicygradienttheorem}
For any joint policy parameter $\bm{\theta}$ and agent $i\in\mathcal{N}$, the gradient of $J(\bm{\theta})$ with respect to $\theta_{i}$ is represented as
\begin{align}\label{thepolicygradienttheorem}
\nabla_{\theta_{i}}J(\bm{\theta})=&\frac{1}{1-\gamma}\mathbb{E}_{\bm{s}\sim d^{\bm{\pi_{\theta}}}_{\bm{\rho}},\bm{a}\sim\bm{\pi_{\theta}}}\Big[\frac{1}{N}\sum_{l\in\mathcal{N}^{\kappa_{p}+1}_{i}}Q^{\bm{\pi_{\theta}}}_{l}(s_{\mathcal{N}_{l}},a_{\mathcal{N}_{l}})\notag\\
&\times\sum_{j\in\mathcal{N}^{\kappa_{p}}_{i}}\nabla_{\theta_{i}}\log\pi_{j}(a_{j}|s_{j},\theta_{j},\theta_{\mathcal{N}^{\kappa_{p}}_{j,-j}})\Big].
\end{align}
\end{theorem}
\par
Theorem~\ref{thelemmaofthepolicygradienttheorem} establishes the expression of the coupled-policy gradient.
It reveals that the coupled-policy gradient $\nabla_{\theta_{i}}J(\bm{\theta})$ of agent $i$ in (\ref{thepolicygradienttheorem}) depends on the local $Q$-functions $\{Q^{\bm{\pi_{\theta}}}_{l}(s_{\mathcal{N}_{l}},a_{\mathcal{N}_{l}})\}_{l\in\mathcal{N}^{\kappa_{p}+1}_{i}}$ from its $(\kappa_{p}+1)$-hop neighbors and the policies $\{\pi_{j}(a_{j}|s_{j},\theta_{j},\theta_{\mathcal{N}^{\kappa_{p}}_{j,-j}})\}_{j\in\mathcal{N}^{\kappa_{p}}_{i}}$ from its $\kappa_{p}$-hop neighbors.
\par
By utilizing $\nabla_{\theta_{i}}J(\bm{\theta})$ in (\ref{thepolicygradienttheorem}), the update of agent $i$'s local policy parameter $\theta_{i,t+1}$ is formally expressed as
\begin{align}\label{thecentralizedupdate}
\theta_{i,t+1}=\theta_{i,t}+\eta_{\theta,t}\nabla_{\theta_{i}}J(\bm{\theta}_{t}),
\end{align}
where {\color{blue}$\bm{\theta}_{t}=(\theta^{\top}_{1,t},\cdots,\theta^{\top}_{N,t})^{\top}$ is the joint policy parameter of agents and} $\eta_{\theta,t}$ is the learning rate of policy parameter in {\color{blue}the} $t$-th iteration of policy parameter.
\begin{remark}
It is important to observe that the iteration (\ref{thecentralizedupdate}) requires each agent $i$ to utilize the exact local $Q$-functions associated with its $(\kappa_{p}+1)$-hop neighbors.
For problems involving high-dimensional state-action spaces,
{\color{blue}estimating} these local $Q$-functions becomes intractable, and the associated complexity grows exponentially with the number of agents.
Although several existing works adopt tabular methods to estimate the $Q$-functions~\cite{QuCLDC2020,QuNIPS2020}, the approximation errors introduced by such estimation inevitably propagate to the policy update, thereby adversely affecting the learning performance.
These limitations motivate the development of a distributed algorithm that avoids the explicit computation of the $Q$-functions or $Q$-tables, thereby enabling scalability in large-scale NMARL problems.
\end{remark}

\section{Distributed scalable coupled policy algorithm}\label{thesectionalgorithmdesign}
To eliminate the need for agents to compute local $Q$-functions and improve the algorithm's scalability, we first introduce a novel concept of the ``neighbors' averaged $Q$-function'' and derive a new expression for coupled-policy gradient.
Based on this derivation, we propose a DSCP algorithm.
\subsection{Neighbors' averaged $Q$-function and a new expression of coupled-policy gradient}
From the definition of the local $Q$-function in (\ref{thelocalQfunctionofglobalpolicy}), we have that the coupled-policy gradient $\nabla_{\theta_{i}}J(\bm{\theta})$ of agent $i$ in (\ref{thepolicygradienttheorem}) depends solely on the state-action pairs within its $(\kappa_{p}+2)$-hop neighbors and the rewards from its $(\kappa_{p}+1)$-hop neighbors.
Motivated by this observation, we introduce a concept of ``neighbors' averaged $Q$-function'' for each agent $i$:
\begin{align}\label{theaction-averagedQfunctionofagenti}
&\widehat{Q^{\bm{\pi_{\theta}}}_{i}}(s_{\mathcal{N}^{\kappa_{p}+2}_{i}},a_{\mathcal{N}^{\kappa_{p}+2}_{i}})\notag\\
=&\mathbb{E}_{\bm{\pi_{\theta}}}\Big[\frac{1}{N}\sum^{\infty}_{t=0}\gamma^{t}\!\!\sum_{j\in\mathcal{N}^{\kappa_{p}+1}_{i}}\!\!r_{j}(s_{\mathcal{N}_{j},t},a_{\mathcal{N}_{j},t})\Big|\notag\\
&s_{\mathcal{N}^{\kappa_{p}+2}_{i},0}=s_{\mathcal{N}^{\kappa_{p}+2}_{i}},a_{\mathcal{N}^{\kappa_{p}+2}_{i},0}=a_{\mathcal{N}^{\kappa_{p}+2}_{i}}\Big].
\end{align}
We emphasize that although the domain of $\widehat{Q^{\bm{\pi_{\theta}}}_{i}}(s_{\mathcal{N}^{\kappa_{p}+2}_{i}},a_{\mathcal{N}^{\kappa_{p}+2}_{i}})$
involves the state-action pairs of the $(\kappa_{p}+2)$-hop neighbors {\color{blue}of} agent $i$, its value can be determined solely by the rewards from its $(\kappa_{p}+1)$-hop neighbors.
This property arises from the specific structure of the local state transition function $\mathcal{P}_{i}(s'_{i}|s_{i},a_{i})$, under which (\ref{theaction-averagedQfunctionofagenti}) remains unaffected by the state-action pairs of agents outside the $(\kappa_{p}+2)$-hop neighborhood of agent $i$.
\par
By using the neighbors' averaged $Q$-function in (\ref{theaction-averagedQfunctionofagenti}), a new expression for coupled-policy gradient {\color{blue}is derived as follows}.
\begin{theorem}\label{thelemmainNNMARLITP}
For any joint policy $\bm{\pi_{\theta}}$, the gradient of $J(\bm{\theta})$ with respect to $\theta_{i}$ {\color{blue}is} represented as
\begin{align}\label{thepolicygradienttheoreminNNMARL-ITP}
\nabla_{\theta_{i}}J(\bm{\theta})
=&\frac{1}{1-\gamma}\mathbb{E}_{\bm{s}\sim d^{\bm{\pi_{\theta}}}_{\bm{\rho}},\bm{a}\sim\bm{\pi_{\theta}}}\Big[\widehat{Q^{\bm{\pi_{\theta}}}_{i}}(s_{\mathcal{N}^{\kappa_{p}+2}_{i}},a_{\mathcal{N}^{\kappa_{p}+2}_{i}})\notag\\
&\times\sum_{j\in\mathcal{N}^{\kappa_{p}}_{i}}\nabla_{\theta_{i}}\log\pi_{j}(a_{j}|s_{j},\theta_{j},\theta_{\mathcal{N}^{\kappa_{p}}_{j,-j}})\Big].
\end{align}
\end{theorem}
\par
Theorem~\ref{thelemmainNNMARLITP} fully exploits the notion of the neighbors' averaged $Q$-function, providing an alternative formulation of the coupled-policy gradient that explicitly incorporates the policies of agent $i$'s $\kappa_{p}$-hop neighbors, along with the term $\widehat{Q^{\bm{\pi_{\theta}}}_{i}}(s_{\mathcal{N}^{\kappa_{p}+2}_{i}},a_{\mathcal{N}^{\kappa_{p}+2}_{i}})$.
{\color{blue}Recall that} each agent $i$ can {\color{blue}estimate} the quantity of $\widehat{Q^{\bm{\pi_{\theta}}}_{i}}(s_{\mathcal{N}^{\kappa_{p}+2}_{i}},a_{\mathcal{N}^{\kappa_{p}+2}_{i}})$ in (\ref{theaction-averagedQfunctionofagenti}) by using reward signals from its $(\kappa_{p}+1)$-hop neighbors.
As a result, the policy gradient $\nabla_{\theta_{i}}J(\bm{\theta})$ in (\ref{thepolicygradienttheoreminNNMARL-ITP}) for each agent $i$ can be computed solely from the state-action pairs and rewards within its limited neighborhood, thus providing a rigorous theoretical basis for the distributed algorithm design presented in the subsequent section.

\subsection{DSCP algorithm}
{\color{blue}Before presenting the DSCP algorithm, we specify the variables maintained by each agent and the information available to it.
Specifically, in the $t$-th iteration of the DSCP algorithm, let $\hat{\theta}^{i}_{j,t}$ denote the local estimate of agent $i$ on the policy parameter of agent $j$, and let $\{\breve{\theta}^{i}_{j,t}\}_{i,j \in \mathcal{N}}$ and $\{p_{i,t}\}_{i \in \mathcal{N}}$ denote the intermediate variables for policy parameter update, with initial values $\breve{\theta}^{i}_{j,1}=\mathbf{0}_{d}$ and $p_{i,1}=1$ for all $i,j\in\mathcal{N}$.
The information collected and exchanged by each agent satisfies the following assumption.
}
{\color{blue}\begin{assumption}\label{theassumptionofcommunication}
In the $t$-th iteration,
let $T_1 \sim \mathrm{Geom}(1-\gamma)$ and $T_2 \sim \mathrm{Geom}(1-\gamma^{1/2})$ be two independent random horizons,
and let $(\bm{s},\bm{a})_{0:T_1+T_2}=(\bm{s}_{0},\bm{a}_{0},\bm{s}_{1},\cdots,\bm{s}_{T_{1}+T_{2}},\bm{a}_{T_{1}+T_{2}})$ be the sample trajectory generated by the executed joint policy. Each agent $i \in \mathcal{N}$ only has access to:
(i) a single snapshot of the state-action pairs from its $\kappa_p$-hop neighbors at time step $T_1$, i.e., $\mathcal{I}^{s,a}_{i,t}=\{ (s_{j,T_1}, a_{j,T_1})| j \in \mathcal{N}_i^{\kappa_p} \}$;
(ii) a trajectory of reward signals from its $(\kappa_p+1)$-hop neighbors over the time interval $[T_1, T_1+T_2]$, i.e., $\mathcal{I}^{r}_{i,t} = \{ r_{j,\tau}| j \in \mathcal{N}_i^{\kappa_p+1},\ T_1 \leq \tau \leq T_1+T_2 \}$;
(iii) the policy parameters and intermediate variables from its direct neighbors, i.e., $\mathcal{I}^{p}_{i,t} = \{\theta_{l,t},\breve{\theta}^{l}_{j,t},p_{l,t}\}_{l \in \mathcal{N}_i,j \in \mathcal{N}}$.
\end{assumption}}
\par
{\color{blue}Assumption~\ref{theassumptionofcommunication} reflects realistic communication constraints~\cite{Yuan2024}, where the information in $\mathcal{I}^{s,a}_{i,t}$ and $\mathcal{I}^{r}_{i,t}$ can be obtained via multi-hop relaying over $\mathcal{G}$, which is consistent with~\cite{QuCLDC2020,QuNIPS2020}.
Specifically, in each iteration, the high-dimensional state-action pairs are sampled only at time step $T_1$, which requires $\kappa_p$ communication rounds for agent $i$ to collect $|\mathcal{N}_i^{\kappa_p}|$ pairs.
The lightweight scalar rewards are sampled at each step in $[T_1,T_1+T_2]$ over a slightly larger neighborhood, each of which requires $\kappa_p+1$ rounds for agent $i$ to collect $|\mathcal{N}_i^{\kappa_p+1}|$ scalars.
The policy parameters are exchanged only once per iteration among direct neighbors, which requires a single round with $\mathcal{O}(|\mathcal{N}_i|Nd)$ scalars per agent.
Note that the scalability of the DSCP algorithm refers to the fact that the environmental information required for policy gradient estimation is restricted to a limited neighborhood, rather than the global state-action information.}
\par
In accordance with Theorem~\ref{thelemmainNNMARLITP} and Assumption~\ref{theassumptionofcommunication}, the design of a distributed algorithm entails addressing two key {\color{blue}challenges}:
\textbf{(i)~Policy execution} and
\textbf{(ii)~Policy update}.
\par
{\color{blue}With respect to policy execution},
under Assumption~\ref{theassumptionofcommunication}, each agent only receives the true policy parameters from its direct neighbors.
When $\kappa_{p} \geq 2$, agents are unable to directly obtain the true policy parameters of all agents within their $\kappa_{p}$-hop neighborhood, which may hinder accurate policy execution.
To address this limitation, we {\color{blue}let} each agent $i$ {\color{blue}maintain} an estimate of the policy parameters of other agents.
Inspired by the push-sum protocol in~\cite{NedicTAC2015}, {\color{blue}we} update $\hat{\theta}^{i}_{j,t}$ {\color{blue}as}
\begin{subequations}\label{thekeyupdateofpolicyparameter}
\begin{numcases}{}
p_{i,t+1}=\sum_{l\in\mathcal{N}_{i}}w_{il}p_{l,t}\label{thekeyupdateofpolicyparameter-2}\\
\hat{\theta}^{i}_{j,t}\!=\!\frac{1}{p_{i,t+1}}\sum_{l\in\mathcal{N}_{i}}w_{il}\breve{\theta}^{l}_{j,t},\label{thekeyupdateofpolicyparameter-3}
\end{numcases}
\end{subequations}
where $p_{i,1}=1$ for all $i \in \mathcal{N}$.
Based on the estimated policy parameters $\{\hat{\theta}^{i}_{j,t}\}_{j \in \mathcal{N}}$,
each agent $i$ executes its local policy as
$\pi_{i}(a_{i} \mid s_{i}, \hat{\theta}^{i}_{i,t},\hat{\theta}^{i}_{\mathcal{N}^{\kappa_{p}}_{i,-i},t})$ during the learning process.
\par
{\color{blue}The challenge of policy update is addressed through the following three steps.}
\par
\textbf{Step 1}: Define $\bm{\hat{\theta}}^{i}_{t}=\big((\hat{\theta}^{i}_{1,t})^{\top},\cdots,(\hat{\theta}^{i}_{N,t})^{\top}\big)^{\top}$, $\bm{\hat{\theta}}_{t}=\big((\bm{\hat{\theta}}^{1}_{t})^{\top},\cdots,(\bm{\hat{\theta}}^{N}_{t})^{\top}\big)^{\top}$, and  $\bm{\hat{\pi}}_{\bm{\hat{\theta}}_{t}}=\prod_{i=1}^{N}\pi_{i}(a_{i}|s_{i},\hat{\theta}^{i}_{i,t},$
$\hat{\theta}^{i}_{\mathcal{N}^{\kappa_{p}}_{i,-i},t})$ as the executed joint policy in {\color{blue}the} $t$-th iteration.
{\color{blue}Let $\nabla_{\theta_i}\tilde{J}(\bm{\hat{\theta}}_t)$ define the surrogate coupled-policy gradient associated with the executed joint policy $\hat{\bm{\pi}}_{\hat{\bm{\theta}}_t}$:}
\begin{align}
\nabla_{\theta_{i}}\tilde{J}(\bm{\hat{\theta}}_{t})=&\frac{1}{1-\gamma}\mathbb{E}_{\bm{s}\sim d^{\bm{\hat{\pi}}_{\bm{\hat{\theta}}_{t}}}_{\bm{\rho}},\bm{a}\sim\bm{\hat{\pi}}_{\bm{\hat{\theta}}_{t}}}\Big[\widehat{Q^{\bm{\hat{\pi}}_{\bm{\hat{\theta}}_{t}}}_{i}}(s_{\mathcal{N}^{\kappa_{p}+2}_{i}},a_{\mathcal{N}^{\kappa_{p}+2}_{i}})\notag\\
&\times\sum_{j\in\mathcal{N}^{\kappa_{p}}_{i}}\nabla_{\theta_{i}}\log\pi_{j}(a_{j}|s_{j},\hat{\theta}^{i}_{j,t},\hat{\theta}^{i}_{\mathcal{N}^{\kappa_{p}}_{j,-j},t})\Big].\label{thepolicygradientestimation1}
\end{align}
Here, $\nabla_{\theta_i}\tilde{J}(\bm{\hat{\theta}}_t)$ is not the true gradient of the objective function under $\bm{\hat{\pi}}_{\bm{\hat{\theta}}_t}$.
Nevertheless, it coincides with the true gradient $\nabla_{\theta_i}J(\bm{\theta}_t)$ when $\bm{\hat{\theta}}_t=\mathbf{1}_N\otimes\bm{\theta}_t$.
To estimate the policy gradient $\nabla_{\theta_{i}}\tilde{J}(\bm{\hat{\theta}}_{t})$ in (\ref{thepolicygradientestimation1}), we employ a geometric 2-horizon sampling method {\color{blue}under Assumption~\ref{theassumptionofcommunication}}.
Specifically, we first {\color{blue}draw} a random horizon $T_{1}\sim\mathrm{Geom}(1-\gamma)$ to generate a finite sample trajectory $(\bm{s},\bm{a})_{0:T_{1}}=(\bm{s}_{0},\bm{a}_{0},\bm{s}_{1},\cdots,\bm{s}_{T_{1}},\bm{a}_{T_{1}})$.
According {\color{blue}to} Assumption~\ref{theassumptionofcommunication}, each agent $i$ can collect $(s_{\mathcal{N}^{\kappa_{p}}_{i},T_{1}},a_{\mathcal{N}^{\kappa_{p}}_{i},T_{1}})$.
Following this, we draw an additional horizon $T_{2}\sim\mathrm{Geom}(1-\gamma^{1/2})$ to generate another finite sample trajectory $(\bm{s},\bm{a})_{T_{1}:T_{1}+T_{2}}=(\bm{s}_{T_{1}},\bm{a}_{T_{1}},\bm{s}_{T_{1}+1},\cdots,\bm{s}_{T_{1}+T_{2}},\bm{a}_{T_{1}+T_{2}})$ for policy gradient estimation, where $T_{2}$ is statistically independent of $T_{1}$.
\par
\textbf{Step 2}:
Each agent $i$ collects a reward trajectory $\{r_{\mathcal{N}^{\kappa_{p}+1}_{i},t}\}_{t=T_{1}}^{T_{1}+T_{2}}$
to estimate the value of {\color{blue}the} neighbors' averaged $Q$-function  $\widehat{Q^{\bm{\hat{\pi}}_{\bm{\hat{\theta}}_{t}}}_{i}}(s_{\mathcal{N}^{\kappa_{p}+2}_{i},T_{1}},a_{\mathcal{N}^{\kappa_{p}+2}_{i},T_{1}})$ in (\ref{theaction-averagedQfunctionofagenti}) as
\begin{align}\label{theestimateofwildehatQi}
\hat{Q}^{\bm{\hat{\pi}}_{\bm{\hat{\theta}}_{t}}}_{i,T_{1}}=\frac{1}{N}\sum_{\tau=0}^{T_{2}}\gamma^{\tau/2}\sum_{j\in\mathcal{N}^{\kappa_{p}+1}_{i}}r_{j,T_{1}+\tau}.
\end{align}
Moreover, each agent $i$ can use the collected sample $(s_{\mathcal{N}^{\kappa_{p}}_{i},T_{1}},a_{\mathcal{N}^{\kappa_{p}}_{i},T_{1}})$
and its local policy parameter estimates $\bm{\hat{\theta}}^{i}_{t}$ to calculate
\begin{align}\label{thegradientestimation}
\hat{g}^{\bm{\hat{\pi}}_{\bm{\hat{\theta}}_{t}}}_{i,T_{1}}\triangleq\sum_{j\in\mathcal{N}^{\kappa_{p}}_{i}}\nabla_{\theta_{i}}\log\pi_{j}(a_{j,T_{1}}|s_{j,T_{1}},\hat{\theta}^{i}_{j,t},\hat{\theta}^{i}_{\mathcal{N}^{\kappa_{p}}_{j,-j},t}).
\end{align}
Based on (\ref{theestimateofwildehatQi}) and (\ref{thegradientestimation}), the coupled-policy gradient estimation for $\nabla_{\theta_{i}}\tilde{J}(\bm{\hat{\theta}}_{t})$ in (\ref{thepolicygradientestimation1}) {\color{blue}is} calculated as
\begin{align}
\hat{\nabla}_{\theta_{i}}\tilde{J}(\bm{\hat{\theta}}_{t})=&\frac{1}{1-\gamma}\hat{Q}^{\bm{\hat{\pi}}_{\bm{\hat{\theta}}_{t}}}_{i,T_{1}}\hat{g}^{\bm{\hat{\pi}}_{\bm{\hat{\theta}}_{t}}}_{i,T_{1}}.\label{thepolicygradientestimation2}
\end{align}
Notably, $\hat{\nabla}_{\theta_{i}}\tilde{J}(\bm{\hat{\theta}}_{t})$ can also be regarded as an approximate of
the true policy gradient $\nabla_{\theta_{i}}J(\bm{\theta}_{t})$ in (\ref{thepolicygradienttheorem}),
based on which the update of $\theta_{i,t+1}$ is formulated as
\begin{align}\label{theupdateoftruepolicyparameters}
\theta_{i,t+1}=\theta_{i,t}+\eta_{\theta,t}\hat{\nabla}_{\theta_{i}}\Tilde{J}(\bm{\hat{\theta}}_{t}),
\end{align}
where $\eta_{\theta,t}$ is the learning rate of policy parameter in {\color{blue}the} $t$-th iteration.
\par
\textbf{Step 3}:
The update of the intermediate variables $\breve{\theta}^{i}_{j,t+1}$ of agent $i$ for $j$ is designed as
\begin{align}
\breve{\theta}^{i}_{j,t+1}=\left\{
\begin{array}{ll}\label{thekeyupdateofpolicyparameter-1}
\sum\limits_{l\in\mathcal{N}_{i}}w_{il}\breve{\theta}^{l}_{j,t},~\mathrm{if}~j\notin\mathcal{N}_{i}\\
\sum\limits_{l\in\mathcal{N}_{i}}w_{il}\breve{\theta}^{l}_{j,t}+w_{ij}N(\theta_{j,t+1}-\theta_{j,t}),\mathrm{o/w}
\end{array}\right.
\end{align}
where
$w_{il}$ is the weight of the edge {\color{blue}$(i,l)$}.
\eqref{thekeyupdateofpolicyparameter-1} {\color{blue}guarantees} the average property, {\color{blue}i.e.,}  $(1/N)\sum_{i=1}^{N}\breve{\theta}^{i}_{j,t}=\theta_{j,t}$ for all $j\in\mathcal{N}$ and $t\geq1$.
{\color{blue}Combined with} (\ref{thekeyupdateofpolicyparameter-1}), the updates in (\ref{thekeyupdateofpolicyparameter}) ensure the average consensus convergence of $\hat{\theta}^{i}_{j,t}$ towards $\theta_{j,t}$.
\par
Combining Steps 1-3, the detailed pseudo-code of the DSCP algorithm is presented in Algorithm~\ref{distributedpolicygradientAlgorithm}.
\begin{algorithm}
\caption{DSCP algorithm}\label{distributedpolicygradientAlgorithm}
\textbf{Input:} {\color{blue}A positive integer} $T\geq2$, the learning rates $\{\eta_{\theta,t}\}^{T}_{t=1}$, the initial policy parameters
$\theta_{j,1}=\breve{\theta}^{i}_{j,1}=\mathbf{0}_{d}$ for all $i,j\in\mathcal{N}$\;
\For{$t=1,2,\cdots,T-1$}{
Reset the initial state $\bm{s}_{0}\sim\bm{\rho}$ and
draw $T_{1}\sim\mathrm{Geom}(1-\gamma)$ and $T_{2}\sim\mathrm{Geom}(1-\gamma^{1/2})$\;
Each agent $i$ computes $\{\hat{\theta}^{i}_{j,t}\}_{j\in\mathcal{N}}$ by (\ref{thekeyupdateofpolicyparameter}) and executes action $a_{i,0}\sim\pi_{i}(\cdot|s_{i,0},\hat{\theta}^{i}_{i,t},\hat{\theta}^{i}_{\mathcal{N}^{\kappa_{p}}_{i,-i},t})$\;
\For{$\tau=1,2,\cdots,T_{1}$}{
Each agent $i$ obtains the local state $s_{i,\tau}$ and executes action $a_{i,\tau}\sim\pi_{i}(\cdot|s_{i,\tau},\hat{\theta}^{i}_{i,t},\hat{\theta}^{i}_{\mathcal{N}^{\kappa_{p}}_{i,-i},t})$\;
}
Each agent $i$ collects
$\{(s_{j,T_{1}},a_{j,T_{1}})\}_{j\in\mathcal{N}^{\kappa_{p}}_{i}}$
from its $\kappa_{p}$-hop neighbors and $\{r_{j,T_{1}}\}_{j\in\mathcal{N}^{\kappa_{p}+1}_{i}}$
from its $(\kappa_{p}+1)$-hop neighbors\;
\For{$\tau=1,2,\cdots,T_{2}$}{
Each agent $i$ obtains the local state $s_{i,T_{1}+\tau}$ and executes action $a_{i,T_{1}+\tau}\!\!\sim\!\!\pi_{i}(\cdot|s_{i,T_{1}+\tau},\!\hat{\theta}^{i}_{i,t},\!\hat{\theta}^{i}_{\mathcal{N}^{\kappa_{p}}_{i,-i},t}\!)$\; Each agent $i$ collects $\{r_{j,T_{1}+\tau}\}_{j\in\mathcal{N}^{\kappa_{p}+1}_{i}}$
from its $(\kappa_{p}+1)$-hop neighbors\;
}
Each agent $i$ calculates  $\hat{\nabla}_{\theta_{i}}\tilde{J}(\bm{\hat{\theta}}_{t})$ by (\ref{thepolicygradientestimation2}), and subsequently updates $\theta_{i,t+1}$ and $\{\breve{\theta}^{i}_{j,t+1}\}_{j\in\mathcal{N}}$ through (\ref{theupdateoftruepolicyparameters}) and (\ref{thekeyupdateofpolicyparameter-1}), respectively\;
}
\textbf{Output:} The joint policy $\bm{\pi}_{\bm{\theta}_{T}}$.
\end{algorithm}
\par
Algorithm~\ref{distributedpolicygradientAlgorithm} entails the collection and exchange of two types of information among agents.
\textbf{(i) Environmental information collection}: each agent $i$ collects state-action pairs $(s_{\mathcal{N}^{\kappa_{p}}_{i},T_{1}},a_{\mathcal{N}^{\kappa_{p}}_{i},T_{1}})$ for its $\kappa_{p}$-hop neighbors and rewards $\{r_{\mathcal{N}^{\kappa_{p}+1}_{i},t}\}_{t=T_{1}}^{T_{1}+T_{2}}$ from its $(\kappa_{p}+1)$-hop neighbors (see Lines~7 and~10).
\textbf{(ii) Parameter information exchange}: each agent $i$ exchanges the true policy parameter $\theta_{i,t}$ and the intermediate variables $\{\breve{\theta}^{i}_{j,t}\}_{j\in\mathcal{N}}$ only with its direct neighbors (see Line~11).
\begin{remark}
{\color{blue}In (\ref{thekeyupdateofpolicyparameter}) and (\ref{thekeyupdateofpolicyparameter-1}), each agent $i$ shares its true policy parameter $\theta_{i,t}$ only with its direct neighbors, while the policy parameters of agents beyond its direct neighborhood are obtained through the intermediate variables $\{\breve{\theta}^{i}_{j,t}\}_{j\in\mathcal{N}}$ exchanged among direct neighbors (see (\ref{thekeyupdateofpolicyparameter-3})).
Note that each individual intermediate variable $\breve{\theta}^{i}_{j,t}$ does not converge to the true parameter $\theta_{j,t}$.
Instead, only their average satisfies $(1/N)\sum_{i=1}^{N}\breve{\theta}^{i}_{j,t}=\theta_{j,t}$, and (\ref{thekeyupdateofpolicyparameter}) ensures that the estimate $\hat{\theta}^{i}_{j,t}$ converges to $\theta_{j,t}$.}
\end{remark}
\begin{remark}
In contrast to $Q$-table-based methods~\cite{QuCLDC2020,QuNIPS2020}, which require each agent to maintain a complete $Q$-table encompassing all possible state-action pairs, the proposed Algorithm~\ref{distributedpolicygradientAlgorithm} obviates the necessity of storing a comprehensive $Q$-table.
Instead, during each iteration, every agent $i$ computes only a single $Q$-value as defined in (\ref{theestimateofwildehatQi}), which is subsequently utilized to construct the policy gradient estimate $\hat{\nabla}_{\theta_{i}}\tilde{J}(\bm{\hat{\theta}}_{t})$ in (\ref{thepolicygradientestimation2}).
Moreover, this estimate serves as
{\color{blue}an unbiased estimator of the surrogate coupled-policy gradient}
$\nabla_{\theta_{i}}\tilde{J}(\bm{\hat{\theta}}_{t})$ under the executed policy {\color{blue}$\bm{\hat{\pi}}_{\bm{\hat{\theta}}_{t}}$}, as formally established in Lemma~\ref{thelemmaofapproximatedpolicygradient}.
\end{remark}
\begin{remark}
It is worth noting that the proposed Algorithm~\ref{distributedpolicygradientAlgorithm} is designed for the case $\kappa_{p}\geq2$.
When $\kappa_{p}=1$, Assumption~\ref{theassumptionofcommunication} implies that each agent can access the true policy parameters of its direct neighbors for executing policy $\bm{\pi}_{\bm{\theta}_{t}}$.
In this scenario, the policy gradient $\nabla_{\theta_{i}}J(\bm{\theta}_{t})$ can be estimated by carrying out Steps~1 and~2 under the executed policy $\bm{\pi}_{\bm{\theta}_{t}}$, without relying on the additional parameter-estimation mechanism required for {\color{blue}the} case where $\kappa_{p}\geq2$.
\end{remark}


\section{Convergence analysis}\label{SectionIV}
In this section, we begin by presenting several standard assumptions established in prior literature~\cite{QuCLDC2020,QuNIPS2020,ZhangSIAM2020,AydinLDCC2023}, followed by the main theoretical results.
\begin{assumption}\label{theassumptionofreward}
For any agent $i\in\mathcal{N}$, there exists a constant $R>0$, such that the instantaneous reward $|r_{i}(s_{\mathcal{N}_{i}},a_{\mathcal{N}_{i}})|\leq R$ for all $(s_{\mathcal{N}_{i}},a_{\mathcal{N}_{i}})\in\mathcal{S}_{\mathcal{N}_{i}}\times\mathcal{A}_{\mathcal{N}_{i}}$.
\end{assumption}
\par
Assumption~\ref{theassumptionofreward} is standard in theoretical analyses of RL and indicates that the instantaneous rewards received by agents are bounded.
\begin{assumption}\label{theassumptionofpolicy}
For any joint policy $\bm{\pi_{\theta}}$, agent $i\in\mathcal{N}$, and agent $j\in\mathcal{N}^{\kappa_{p}}_{i}$, $\nabla_{\theta_{j}}\log\pi_{i}(a_{i}|s_{i},\theta_{i},\theta_{\mathcal{N}^{\kappa_{p}}_{i,-i}})$ exists and satisfies $\|\nabla_{\theta_{j}}\log\pi_{i}(a_{i}|s_{i},\theta_{i},\theta_{\mathcal{N}^{\kappa_{p}}_{i,-i}})\|\leq B$ with $B>0$ for any  $(s_{i},a_{i})\in\mathcal{S}_{i}\times\mathcal{A}_{i}$.
Moreover, $\nabla_{\theta_j}\log\pi_i(a_i|s_i,\theta_i,\theta_{\mathcal{N}^{\kappa_p}_{i,-i}})$ is $L$-Lipschitz continuous, i.e., for any $\bm{\theta},\bm{\theta}'\in\mathbb{R}^{dN}$ and $(s_i,a_i)\in\mathcal{S}_i\times\mathcal{A}_i$,
{\color{blue}$\|\nabla_{\theta_{j}}\log\pi_{i}(a_{i}|s_{i},\theta_{i},\theta_{\mathcal{N}^{\kappa_{p}}_{i,-i}})-\nabla_{\theta_{j}}\log\pi_{i}(a_{i}|s_{i},\theta'_{i},\theta'_{\mathcal{N}^{\kappa_{p}}_{i,-i}})\|\leq L\|\bm{\theta}-\bm{\theta}'\|$.}
\end{assumption}
\par
{\color{blue}Assumption~4 implies that, for each agent $i\in\mathcal{N}$, all existing cross-score terms $\nabla_{\theta_j}\log\pi_i(a_i|s_i,\theta_i,\theta_{\mathcal{N}^{\kappa_p}_{i,-i}})$ with $j\in\mathcal{N}^{\kappa_p}_i$ are uniformly bounded by $B$ and uniformly $L$-Lipschitz.
In particular, this condition is satisfied by, e.g., softmax or log-linear policy classes with finite action spaces and bounded features, including the coupled softmax policy used in Section~\ref{SectionSimulations}.}
\begin{assumption}\label{theassumptionofnetwork}
The learning rate in Algorithm~\ref{distributedpolicygradientAlgorithm} {\color{blue}satisfies $\eta_{\theta,t}=\frac{c_\theta}{t}$ for all $t\geq1$, where $c_\theta>0$ is a constant.}
\end{assumption}
\par
Assumption~\ref{theassumptionofnetwork} {\color{blue}is standard} in distributed optimization field and can ensure the convergence of estimated policy parameters to their true values.
\subsection{Intermediate lemmas}\label{subsectionIntermediatelemmas}
\begin{lemma}\cite{NedicTAC2015}\label{thelemmaoftime-varyingnetwork}
Suppose Assumption~\ref{theassumptionontheunderlyingnetwork} holds.
For each integer $k\geq1$, there is a stochastic vector sequence $\{\phi_{t}\}_{t\geq1}$ such that for all $i,j\in\mathcal{N}$ and $t\geq k\geq1$,
\begin{align}
|W^{t-k+1}_{ij}-\phi_{i,t}|\leq M_{1}\lambda^{t-k},
\end{align}
where $M_{1}>0$, $\lambda\in(0,1)$, and $W^{t-k+1}_{ij}$
represents the element located in the $i$-th row and the $j$-th column of $W^{t-k+1}$, and $\phi_{i,t}$ denotes the $i$-th element in $\phi_{t}$.
\end{lemma}
\par
This {\color{blue}lemma} can be derived from Corollary~2 in~\cite{NedicTAC2015}.
Moreover,
{\color{blue}it also follows from Lemma~\ref{thelemmaoftime-varyingnetwork} that}
there {\color{blue}exist $w_{max}>1$ and $\kappa>0$, such that $\|W^{t-k+1}\|\leq w_{max}$ and $N\phi_{i,t}+(W^{t}-\phi_{t}\mathbf{1}^{\top}_{N})_{i}\mathbf{1}\geq\kappa$ for all $t\geq k\geq1$, where $(W^{t}-\phi_{t}\mathbf{1}^{\top}_{N})_{i}$ denotes the $i$-th row of the matrix $W^{t}-\phi_{t}\mathbf{1}^{\top}_{N}$.}
\par
For notational convenience, for any $\kappa_p\geq1$,
we define
\begin{align}\label{themaxnumberofkappap}
M_{\kappa_{p}}=\max_{i\in\mathcal{N}}|\mathcal{N}^{\kappa_{p}}_{i}|,
\end{align}
which characterizes the largest $\kappa_p$-hop neighborhood size among all agents.
Recalling the definition of $\kappa_{p}$-hop neighborhoods in Section~\ref{theintroduvtionofnetwork}, it follows that $1\leq M_{\kappa_p}\leq N$ for all $\kappa_{p}\geq1$ and is non-decreasing with respect to $\kappa_p$.
\par
{\color{blue}Let $\mathcal{F}_t$ denote the $\sigma$-algebra generated by all random horizons and sample trajectories drawn before the $t$-th iteration of Algorithm~\ref{distributedpolicygradientAlgorithm}.}
{\color{blue}In the $t$-th iteration, the initial state is reset as $\bm{s}_0\sim\bm{\rho}$, and $T_1,T_2$ and the sample trajectory are newly generated under $\bm{\hat{\pi}}_{\hat{\bm{\theta}}_t}$, independently of the past given $\mathcal{F}_t$.}
Based on the above Assumptions~\ref{theassumptionofreward}-\ref{theassumptionofpolicy}, we have the following lemmas.
\begin{lemma}\label{thelemmaofapproximatedpolicygradient}
Suppose Assumptions~\ref{theassumptionofreward}-\ref{theassumptionofpolicy} hold.
For the estimated policy parameter sequence $\{\bm{\hat{\theta}}_{t}\}_{1\leq t\leq T}$ generated by Algorithm~\ref{distributedpolicygradientAlgorithm}, we have\\
(i) The estimated coupled-policy gradient
$\hat{\nabla}_{\theta_{i}}\Tilde{J}(\bm{\hat{\theta}}_{t})$ is bounded, i.e.,  $\|\hat{\nabla}_{\theta_{i}}\Tilde{J}(\bm{\hat{\theta}}_{t})\|\leq\hat{L}$ with $\hat{L}=\frac{BRM_{\kappa_{p}}M_{\kappa_{p}+1}}{(1-\gamma)(1-\gamma^{1/2})N}$;\\
(ii) $\hat{\nabla}_{\theta_{i}}\Tilde{J}(\bm{\hat{\theta}}_{t})$ is an unbiased estimate of $\nabla_{\theta_{i}}\Tilde{J}(\bm{\hat{\theta}}_{t})$, i.e., $\mathbb{E}[\hat{\nabla}_{\theta_{i}}\Tilde{J}(\bm{\hat{\theta}}_{t})|\mathcal{F}_{t}]=\nabla_{\theta_{i}}\Tilde{J}(\bm{\hat{\theta}}_{t})$, {\color{blue}where the conditional expectation is taken over the random horizons $T_1,T_2$ and the sample trajectory in the $t$-th iteration.}
\end{lemma}
\par
The first part establishes that the norm of the estimated policy gradient $\hat{\nabla}_{\theta_{i}}\tilde{J}(\bm{\hat{\theta}}_{t})$ is upper-bounded, with the bound explicitly dependent on the choice of $\kappa_{p}$.
Moreover, the second part demonstrates that Algorithm~\ref{distributedpolicygradientAlgorithm} provides each agent $i$ with an unbiased estimate of {\color{blue}the surrogate coupled-policy gradient under} the executed policy $\bm{\hat{\pi}}_{\bm{\hat{\theta}}_{t}}$ during the $t$-th iteration, which is essential for establishing the convergence of Algorithm~\ref{distributedpolicygradientAlgorithm} in the subsequent Theorem~\ref{thetheoremconvergenceofpolicygradient}.
\par
{\color{blue}With} $L_{1}=\big(\frac{LRM_{\kappa_{p}}M_{\kappa_{p}+1}}{(1-\gamma)^{2}N}+\frac{(1+\gamma)B^{2}RM^{2}_{\kappa_{p}}M_{\kappa_{p}+1}}{(1-\gamma)^{3}N^{\frac{1}{2}}}\big)$ and $L_{2}=\big(\frac{LRM_{\kappa_{p}}M_{\kappa_{p}+1}}{(1-\gamma)^{2}N^{\frac{1}{2}}}+\frac{(1+\gamma)B^{2}RM^{2}_{\kappa_{p}}M_{\kappa_{p}+1}}{(1-\gamma)^{3}}\big)$, we have the following results.
\begin{lemma}\label{thesmoothofpolicygradient}
Suppose Assumptions~\ref{theassumptionofreward}-\ref{theassumptionofpolicy}  hold.
For any agent $i\in\mathcal{N}$ and joint policy parameters  $\bm{\theta},\bm{\theta}'\in\mathbb{R}^{dN}$, we have
\begin{align}
\|\nabla_{\theta_{i}}J(\bm{\theta})-\nabla_{\theta_{i}}J(\bm{\theta}')\|\leq L_{1}\|\bm{\theta}-\bm{\theta}'\|,
\end{align}
which also means that $J(\bm{\theta})$ is $L_{2}$-smooth.
\end{lemma}
\par
Lemma~\ref{thesmoothofpolicygradient} proves that under the Assumptions~\ref{theassumptionofreward}-\ref{theassumptionofpolicy}, the objective function $J(\bm{\theta})$ with coupled policy in (\ref{theobjectivefunction}) {\color{blue}is} $L_{2}$-smooth, with the constant $L_2$ depending explicitly on $\kappa_p$.
\par
Similar to Lemma~\ref{thesmoothofpolicygradient}, the surrogate coupled-policy gradient $\nabla_{\theta_{i}}\tilde{J}(\bm{\hat{\theta}})$ in (\ref{thepolicygradientestimation1}) under the executed joint policy $\bm{\hat{\pi}}_{\bm{\hat{\theta}}}$ satisfies the following property.
\begin{corollary}\label{theerrorbetweenpolicygradient}
Suppose Assumptions~\ref{theassumptionofreward}-\ref{theassumptionofpolicy}  hold.
For any agent $i\in\mathcal{N}$, {\color{blue}and} any $\bm{\theta}\in\mathbb{R}^{dN}$ and $\bm{\hat{\theta}}\in\mathbb{R}^{dN^{2}}$,  {\color{blue}the surrogate coupled-policy gradient under $\bm{\hat{\pi}_{\hat{\theta}}}$ defined in (\ref{thepolicygradientestimation1}) satisfies}
$\nabla_{\theta_{i}}J(\bm{\theta})=\nabla_{\theta_{i}}\tilde{J}(\mathbf{1}_{N}\otimes\bm{\theta})$ and
\begin{align}
\|\nabla_{\theta_{i}}\tilde{J}(\mathbf{1}_{N}\otimes\bm{\theta})-\nabla_{\theta_{i}}\tilde{J}(\bm{\hat{\theta}})\|\leq L_{1}\|\mathbf{1}_{N}\otimes\bm{\theta}-\bm{\hat{\theta}}\|.\label{resultofcorollary1}
\end{align}
\end{corollary}
\par
Corollary~\ref{theerrorbetweenpolicygradient} closely parallels Lemma~\ref{thesmoothofpolicygradient}.
It characterizes the error between {\color{blue}the surrogate coupled-policy gradient} under the executed joint policy $\bm{\hat{\pi}}_{\hat{\bm{\theta}}}$ and the true policy gradient $\nabla_{\theta_{i}}J(\bm{\theta})$.

\subsection{Main results}\label{themainresults}
The main results, such as the convergence of the policy parameter estimation and {\color{blue}the asymptotic} first-order stationarity
convergence of Algorithm~\ref{distributedpolicygradientAlgorithm}, are presented as below.
\begin{theorem}\label{thelemmaofpolicyparameterconvergence}
Suppose Assumptions~\ref{theassumptionontheunderlyingnetwork}-\ref{theassumptionofnetwork} hold.
{\color{blue}Let $\Psi_{\theta}=\frac{\sqrt{Nd}(1+w_{max})w_{max}c_{\theta}BRM_{1}}{(1-\lambda)(1-\gamma)(1-\gamma^{\frac{1}{2}})\kappa}$.}
In Algorithm~\ref{distributedpolicygradientAlgorithm}, for all {\color{blue}$t\geq2$}, $i,j\in\mathcal{N}$, the local policy parameter estimate $\hat{\theta}^{i}_{j,t}$ of agent $i$ with respect to agent $j$ satisfies
\begin{align}
\|\hat{\theta}^{i}_{j,t}-\theta_{j,t}\|\leq&{\color{blue}\frac{\Psi_{\theta}M_{\kappa_{p}}M_{\kappa_{p}+1}(1+\log{t})}{t-1}}\label{theconvergenceofpolicyparameter}
\end{align}
and
\begin{align}
{\color{blue}\lim_{t\rightarrow\infty}\|\hat{\theta}^{i}_{j,t}-\theta_{j,t}\|=0,\forall i,j\in\mathcal{N}.}\label{theconvergenceofpolicyparameter2}
\end{align}
\end{theorem}
\par
{\color{blue}Theorem~\ref{thelemmaofpolicyparameterconvergence} establishes that $\hat{\theta}^{i}_{j,t}$ converges to $\theta_{j,t}$ as $t\rightarrow\infty$ for any realization of the sampled trajectories, without relying on any expectation, since the estimated policy gradient is uniformly bounded (see~(i) of Lemma~\ref{thelemmaofapproximatedpolicygradient}) and decaying learning rate $\eta_{\theta,t}=\frac{c_\theta}{t}$.
Moreover, the constant $\Psi_\theta$ depends only on the network topology (i.e., $w_{\max}$, $\kappa$, $\lambda$ and $M_1$), the problem parameters (i.e., $R$, $B$, $\gamma$, $N$ and $d$) and the learning rate (i.e., $c_\theta$),
while the effect of the coupled-policy radius $\kappa_p$ on the error bound is explicitly characterized by $M_{\kappa_p}M_{\kappa_p+1}$.
This theorem ensures that the agents can generate reliable samples based on the executed joint policy $\bm{\hat{\pi}}_{\bm{\hat{\theta}}_{t}}$ for cooperative policy optimization.
}

\begin{theorem}\label{thetheoremconvergenceofpolicygradient}
Suppose Assumptions~\ref{theassumptionontheunderlyingnetwork}-\ref{theassumptionofnetwork} hold.
There exists a positive constant $J_{\mathrm{sup}}(M_{\kappa_{p}},M_{\kappa_{p}+1})>0$, which depends on the neighborhood sizes $M_{\kappa_{p}}$ and $M_{\kappa_{p}+1}$ and is monotonically increasing in both arguments, such that the policy parameter sequence $\{\bm{\theta}_{t}\}_{t\geq1}$ generated by Algorithm~\ref{distributedpolicygradientAlgorithm} {\color{blue}satisfies
$\mathbb{E}\big[\sum_{t=1}^{\infty}\eta_{\theta,t}\|\nabla_{\bm{\theta}}J(\bm{\theta}_{t})\|^{2}\big]\leq J_{\mathrm{sup}}(M_{\kappa_{p}},M_{\kappa_{p}+1})$,
\begin{align}
\lim_{t\rightarrow\infty}\|\nabla_{\bm{\theta}}J(\bm{\theta}_{t})\|=0\quad\text{almost surely},\label{thelastinequalityintheorembefore1}
\end{align}
and
\begin{align}
\lim_{t\rightarrow\infty}\mathbb{E}\big[\|\nabla_{\bm{\theta}}J(\bm{\theta}_{t})\|\big]=0.\label{thelastinequalityintheorem}
\end{align}}
\end{theorem}
\par
{\color{blue}Theorem~\ref{thetheoremconvergenceofpolicygradient} shows that the norm of $\nabla_{\bm{\theta}}J(\bm{\theta}_t)$ converges to zero both almost surely and in expectation, which indicates that Algorithm~\ref{distributedpolicygradientAlgorithm} is asymptotically first-order stationary.}

\subsection{Extended results for reward function with larger coupled radius}
The proposed framework can be extended to settings where rewards depend on larger neighborhood scopes, as stated in the following assumption.
\begin{assumption}\label{theassumptionofreward2}
The reward function of agent $i$ adopts the form $r_{i}(s_{\mathcal{N}^{\kappa_r}_i},a_{\mathcal{N}^{\kappa_r}_i})$, where $1\leq\kappa_r \leq \kappa_p$ represents a coupled radius that depicts the reward dependency range.
For any agent $i\in\mathcal{N}$, there exists a constant $R>0$, such that the instantaneous reward $|r_{i}(s_{\mathcal{N}^{\kappa_{r}}_{i}},a_{\mathcal{N}^{\kappa_{r}}_{i}})|\leq R$ for all $(s_{\mathcal{N}^{\kappa_{r}}_{i}},a_{\mathcal{N}^{\kappa_{r}}_{i}})\in\mathcal{S}_{\mathcal{N}^{\kappa_{r}}_{i}}\times\mathcal{A}_{\mathcal{N}^{\kappa_{r}}_{i}}$.
\end{assumption}
\par
It should be noted that the restriction $\kappa_{p}\geq\kappa_{r}$ in Assumption~\ref{theassumptionofreward2} is employed to guarantee that the coupled policy can comprehensively capture the inter-agent reward dependencies.
\par
Under Assumption~\ref{theassumptionofreward2}, the main results in Section~\ref{themainresults} remain valid after appropriately adjusting the locality radii: $(\kappa_p+1)\to(\kappa_p+\kappa_r)$ in (\ref{thepolicygradienttheorem}), (\ref{theaction-averagedQfunctionofagenti}), (\ref{theestimateofwildehatQi}), and $(\kappa_p+2)\to(\kappa_p+2\kappa_r)$ in (\ref{theaction-averagedQfunctionofagenti}), (\ref{thepolicygradienttheoreminNNMARL-ITP}), (\ref{thepolicygradientestimation1}).
Moreover, Algorithm~\ref{distributedpolicygradientAlgorithm} can then be similarly redeveloped under {\color{blue}the} following assumption.
\begin{assumption}\label{themodifiedassumptionofcommunication}
{\color{blue}In each policy-update iteration, Assumption~\ref{theassumptionofcommunication} holds with the reward information enlarged from
the $(\kappa_p+1)$-hop to the $(\kappa_p+\kappa_r)$-hop neighborhood, i.e.,
$\mathcal{I}^r_{i,t}=\{r_{j,\tau}\,|\,j\in\mathcal{N}_i^{\kappa_p+\kappa_r},
\,T_1\le\tau\le T_1+T_2\}$.}
\end{assumption}
\par
Based on Assumption~\ref{themodifiedassumptionofcommunication}, Algorithm~\ref{distributedpolicygradientAlgorithm} can be extended in a direct manner, yielding the following results.
\begin{corollary}\label{corollaryfinal}
Suppose Assumptions~\ref{theassumptionontheunderlyingnetwork} and~\ref{theassumptionofpolicy}-\ref{themodifiedassumptionofcommunication} hold.
In the extended Algorithm~\ref{distributedpolicygradientAlgorithm} for $\{r_{i}(s_{\mathcal{N}^{\kappa_{r}}_{i}},a_{\mathcal{N}^{\kappa_{r}}_{i}})\}_{i\in\mathcal{N}}$, the local policy parameter estimate $\hat{\theta}^{i}_{j,t}$ of agent $i$ with respect to agent $j$
converges to agent $j$'s true policy parameter $\theta_{j,t}$ {\color{blue}for every realization of the sampled trajectories}
with a rate $\mathcal{O}(\frac{M_{\kappa_{p}}M_{\kappa_{p}+\kappa_{r}}\log{t}}{t})$.
Moreover, there exists a positive constant $J_{\mathrm{sup}}(M_{\kappa_{p}},M_{\kappa_{p}+\kappa_{r}})>0$ such that {\color{blue} Theorem~\ref{thetheoremconvergenceofpolicygradient} holds with $J_{\mathrm{sup}}(M_{\kappa_{p}},M_{\kappa_{p}+1})$ replaced by $J_{\mathrm{sup}}(M_{\kappa_{p}},M_{\kappa_{p}+\kappa_{r}})$.
}
\end{corollary}
\par
Corollary~\ref{corollaryfinal} directly extends Theorems~\ref{thelemmaofpolicyparameterconvergence} and~\ref{thetheoremconvergenceofpolicygradient}, demonstrating our DSCP algorithm's generalizability to different coupled ranges in the reward function.

\section{Simulation}\label{SectionSimulations}
This section evaluates Algorithm~\ref{distributedpolicygradientAlgorithm} in a robot swarm path-planning environment.
Although a related setup appears in~\cite{DaiTAC2025}, the task considered here is significantly more complex.


\subsection{The description of the path planning problem}
\begin{figure}[!htb]
\centering
\includegraphics[width=0.7\hsize]{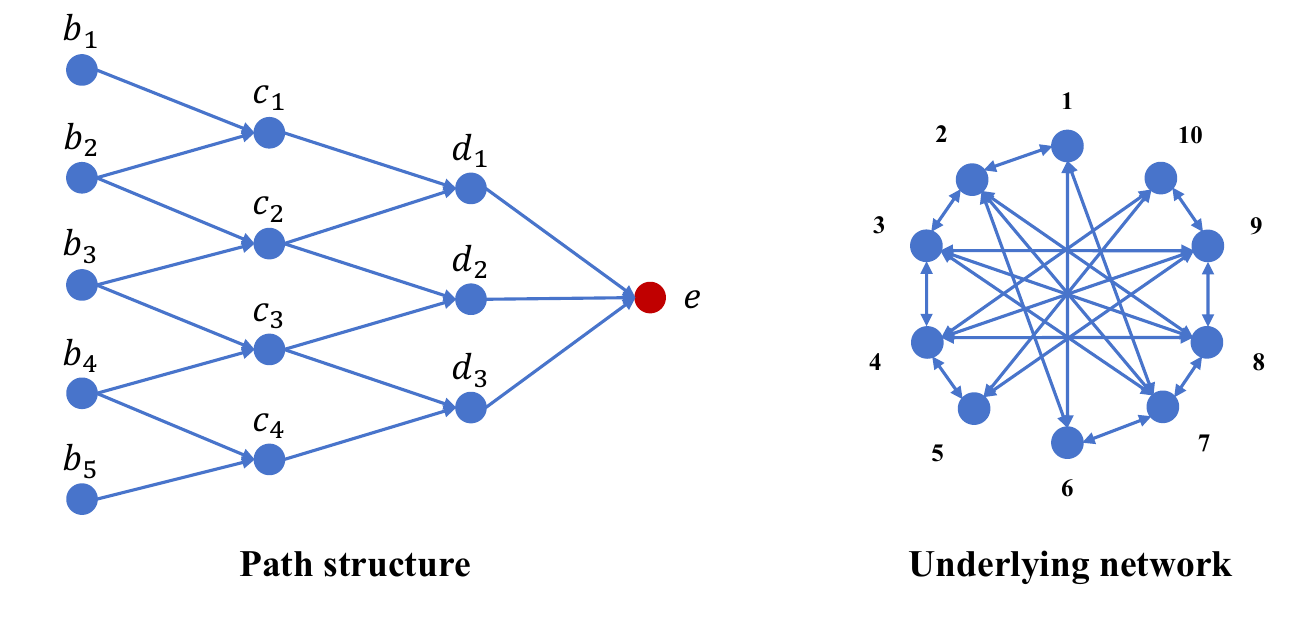}
\caption{The left panel depicts the path structure with~13 locations, while the right panel shows the underlying network among~10 agents.}\label{Pathstructureandunderlyingnetwork}
\end{figure}

The path planning problem of $N=10$ robots (i.e., agents) on the acyclic path structure (as shown in Fig.~\ref{Pathstructureandunderlyingnetwork}) is considered, where the ``blue'' nodes $\{b_{1},b_{2},\cdots,b_{5}\}$ represent the set of starting locations for agents and the
 ``red'' node represents the destination location.
\par
We formulate {\color{blue}an} NMARL model $\big(\mathcal{G}(\mathcal{N},\mathcal{E}),\{\mathcal{S}_{i}\}_{i\in\mathcal{N}},$
$\{\mathcal{A}_{i}\}_{i\in\mathcal{N}},\{\mathcal{P}^{s}_{i}\}_{i\in\mathcal{N}},\{r_{i}\}_{i\in\mathcal{N}},\gamma\big)$ for the path planning problem, where $\mathcal{G}$ denotes the underlying network among agents with $\mathcal{N}=\{1,\cdots,N\}$ representing the set of all agents.
The detailed structure of the underlying network is elaborated in Fig.~\ref{Pathstructureandunderlyingnetwork}.
$\mathcal{S}_{i}=\{b_{1},b_{2},b_{3},b_{4},b_{5},c_{1},c_{2},c_{3},c_{4},d_{1},d_{2},d_{3},e\}$ denotes the local state space of agent $i\in\mathcal{N}$, and $\mathcal{A}_{i}=\{0,1,2\}$ denotes its local action space.
In this problem, the state transition $s'_{i}=\mathcal{P}^{s}_{i}(s_{i},a_{i})$ of agent $i$ is governed by the following rules:
(i) The local action ``0'' means that agent $i$ will remain stationary at the current location for one step;
(ii) If the out-degree of the current location is 2, the actions ``1'' and ``2'' indicate that agent $i$ will follow the upper edge and the lower edge, respectively;
(iii) If the action selected by agent $i$ exceeds the {\color{blue}out-degree} of the current location, it will remain at the current location for one step.
{\color{blue}Letting}  $P^{s_{i},a_{i}}_{i}=\big(s_{i}\rightarrow\mathcal{P}^{s}_{i}(s_{i},a_{i})\big)$ denote the movement of agent $i$, {\color{blue}we define} the reward function $r_{i}(s_{\mathcal{N}_{i}},a_{\mathcal{N}_{i}})$ of agent $i$ as
\begin{align}\notag
r_{i}(s_{\mathcal{N}_{i}},a_{\mathcal{N}_{i}})=\left\{
\begin{array}{ll}
{\color{blue}0, ~~~~\mathrm{if}~s_i=e}\\
-r_{\epsilon},~\mathrm{if}~s_{i}=\mathcal{P}^{s}_{i}(s_{i},a_{i})\\
-r_{\epsilon}-\frac{0.5|\{j\in\mathcal{N}_{i}|P^{s_{i},a_{i}}_{i}=P^{s_{j},a_{j}}_{j}\}|}{N},\mathrm{o/w},
\end{array}\right.
\end{align}
where the first term $r_{\epsilon} = 0.5$ denotes the time cost incurred at each step, while the second term captures the collision penalty when agent $i$ shares the same path with its neighbors.
Moreover, $|\{ j \in \mathcal{N}_i \mid P^{s_i,a_i}_i = P^{s_j,a_j}_j\}|$ represents the number of neighbor agents that share the same path with agent $i$.
\par
In this path planning problem, we set the initial locations of agents as $\{b_{1},b_{2},b_{3},b_{4},b_{5},b_{1},b_{2},b_{3},b_{4},b_{5}\}$ and the discount factor as $\gamma=0.9$.
The objective of agents is to reach the destination as quickly as possible while minimizing collisions, i.e., maximize the objective function in (\ref{theobjectivefunction}).
\subsection{The results of Algorithm~\ref{distributedpolicygradientAlgorithm}}
In Algorithm~\ref{distributedpolicygradientAlgorithm},
the local coupled policy $\pi_{i}(a_{i}|s_{i},\theta_{i},\theta_{\mathcal{N}^{\kappa_{p}}_{i,-i}})$ for agent $i$ is defined as
\begin{align}
&\pi_{i}(a_{i}|s_{i},\theta_{i},\theta_{\mathcal{N}^{\kappa_{p}}_{i,-i}})\notag\\
=&\frac{\exp\big(0.9\theta_{i,s_{i},a_{i}}+(0.1/|\mathcal{N}^{\kappa_{p}}_{i,-i}|)\sum_{j\in\mathcal{N}^{\kappa_{p}}_{i,-i}}\theta_{j,s_{i},a_{i}}\big)}{\sum_{a'_{i}}\exp\big(0.9\theta_{i,s_{i},a'_{i}}+(0.1/|\mathcal{N}^{\kappa_{p}}_{i,-i}|)\sum_{j\in\mathcal{N}^{\kappa_{p}}_{i,-i}}\theta_{j,s_{i},a'_{i}}\big)},\notag
\end{align}
where $\theta_{i}\in\mathbb{R}^{|\mathcal{S}_{i}||\mathcal{A}_{i}|}$ for all $i\in\mathcal{N}$.
\subsubsection{Performance of Algorithm~\ref{distributedpolicygradientAlgorithm}}
To demonstrate the advantages of our Algorithm~\ref{distributedpolicygradientAlgorithm}, we {\color{blue}compare} it with the scalable actor-critic (SAC) algorithm~\cite{QuCLDC2020} and distributed policy gradient (DPG) algorithm~\cite{LiuarXiv2023} under multiple random seeds.
\begin{figure}[!htb]
\centering
\includegraphics[width=0.6\hsize]{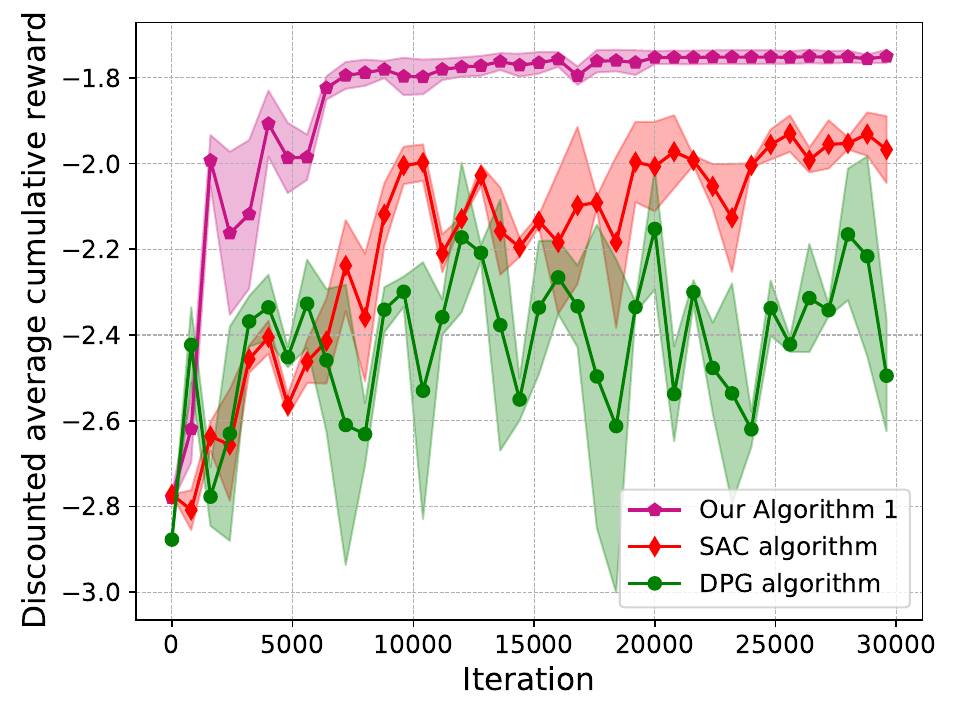}
\caption{The evolution of $J(\bm{\theta}_{t})$ produced by Algorithm~\ref{distributedpolicygradientAlgorithm}, SAC algorithm, and DPG algorithm.
}\label{objectivenetwork2}
\end{figure}
\par
The evolution of objective function $J(\bm{\theta}_{t})$ produced by Algorithm~\ref{distributedpolicygradientAlgorithm} with $\kappa_{p}=2$, SAC algorithm, and DPG algorithm are shown in Fig.~\ref{objectivenetwork2}, where the solid line represents the average value of $J(\bm{\theta}_{t})$ over multiple random seeds and the shaded area represents its standard deviation.
As shown in Fig.~\ref{objectivenetwork2}, Algorithm~\ref{distributedpolicygradientAlgorithm} achieves a higher discounted average cumulative reward than both the SAC and DPG algorithms in the tested setting.
These results demonstrate that the coupled policy framework and {\color{blue}the distributed update scheme} provide enhanced learning capabilities compared to the independent policy approaches in SAC algorithm and DPG algorithm.
We also observe that the DPG algorithm converges more slowly, which can be attributed to the fact that it directly estimates the state-value function rather than the $Q$-function.

\begin{figure}[htbp]
    \centering
    \subfigure[$\frac{1}{N^{2}}\sum_{i=1}^{N}\sum_{j=1}^{N}\|\hat{\theta}^{i}_{j,t}-\theta_{j,t}\|$]{\includegraphics[width=0.24\textwidth]{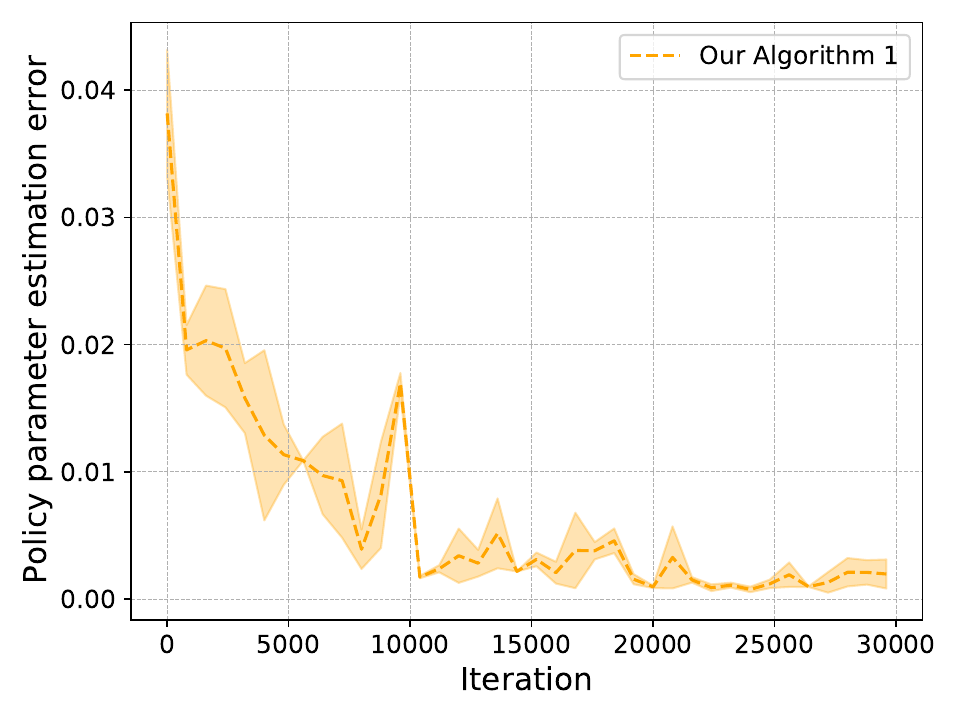}\label{errornetwork2}}
    \subfigure[$\|\nabla_{\bm{\theta}}J(\bm{\theta}_{t})\|$]{\includegraphics[width=0.24\textwidth]{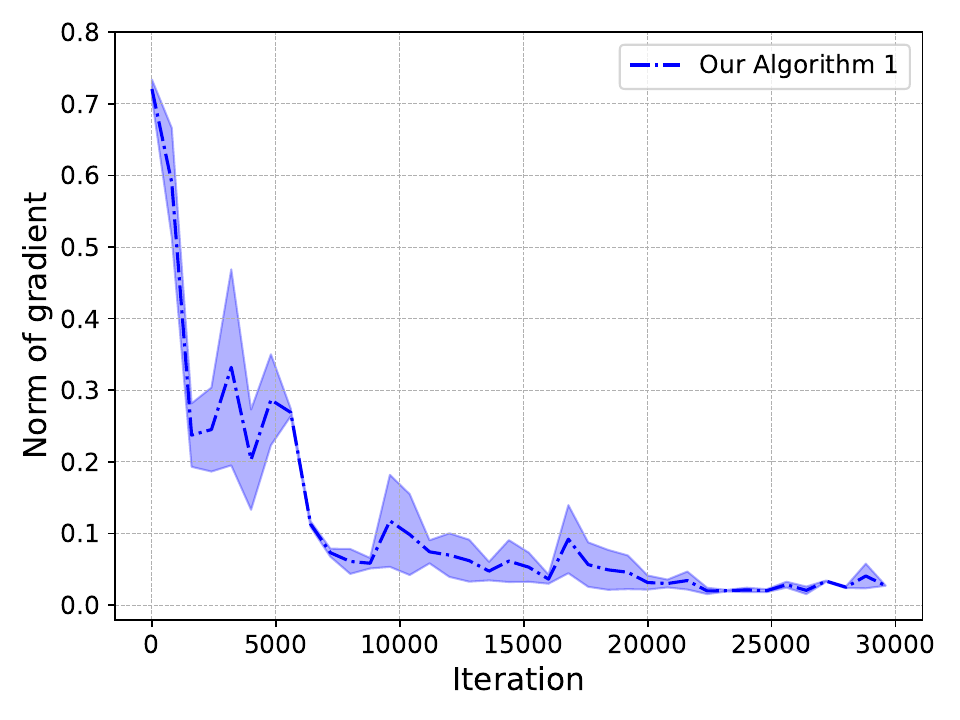}\label{gradientnetwork2}}
    \caption{The evolution of the policy parameter estimation error and the norm of policy gradient produced by Algorithm~\ref{distributedpolicygradientAlgorithm}.}
    \label{totalerror}
\end{figure}

\par
Fig.~\ref{totalerror} shows the evolution of the policy parameter estimation error $(1/N^{2})\sum_{i=1}^{N}\sum_{j=1}^{N}\|\hat{\theta}^{i}_{j,t}-\theta_{j,t}\|$ and the norm of policy gradient $\|\nabla_{\bm{\theta}}J(\bm{\theta}_{t})\|$ produced by Algorithm~\ref{distributedpolicygradientAlgorithm}.
As illustrated in Fig.~\ref{errornetwork2},
the policy parameter estimation error
{\color{blue}decreases to near zero, which is consistent with} the convergence
of $\hat{\theta}^{i}_{j,t}$ to $\theta_{j,t}$ established in  Theorem~\ref{thelemmaofpolicyparameterconvergence}.
Furthermore, Fig.~\ref{gradientnetwork2} shows that the norm of the policy gradient, $\|\nabla_{\bm{\theta}}J(\bm{\theta}_{t})\|$,
{\color{blue}also decreases to near zero during the learning process, which is consistent with the asymptotic first-order stationarity} of Algorithm~\ref{distributedpolicygradientAlgorithm} established in Theorem~\ref{thetheoremconvergenceofpolicygradient}.

\subsubsection{Ablation experiment}
An ablation study is conducted by setting $\kappa_p=0,1,2$ in Algorithm~\ref{distributedpolicygradientAlgorithm}.
Since the coupled policy in Section~\ref{ModelofNMARLproblem} is defined for $\kappa_p\geq1$, we define the case $\kappa_p=0$ separately as the independent softmax policy
\begin{align}
\pi_i(a_i|s_i,\theta_i)=\frac{\exp(\theta_{i,s_i,a_i})}{\sum_{a'_i\in\mathcal{A}_i}\exp(\theta_{i,s_i,a'_i})}.\notag
\end{align}
{\color{blue}In this case, Algorithm~\ref{distributedpolicygradientAlgorithm} requires neither policy parameter estimation nor neighbors' state-action pairs, and each agent only uses the rewards from its 1-hop neighbors.}
\begin{figure}[!htb]
\centering
\includegraphics[width=0.6\hsize]{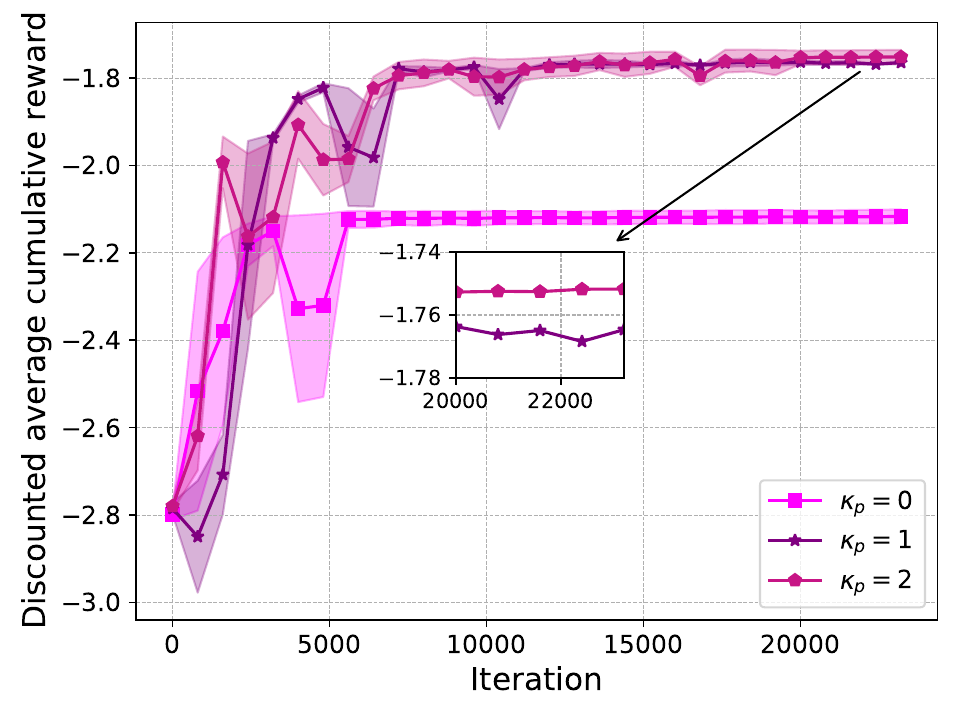}
\caption{The performances of  Algorithm~\ref{distributedpolicygradientAlgorithm} with different $\kappa_{p}$ on objective function.}\label{fig:ablation}
\end{figure}
\par
\textbf{{\color{blue}Empirical convergence speed}}:
As observed in Fig.~\ref{fig:ablation}, when $\kappa_p=0$, the curve reaches a stable level the earliest, at approximately $5.6\times10^3$ iterations.
{\color{blue}However, the faster stabilization is accompanied by noticeably degraded solution quality, since the policy with $\kappa_p=0$ is learned independently without exploiting the neighbors' policy parameters or state-action pairs.}
For $\kappa_p=1$ and $\kappa_p=2$, {\color{blue}the curves} reach a stable level at approximately $1.2\times10^4$ and $2.0\times10^4$ iterations, respectively.
{\color{blue}That is, in this simulation, the convergence of Algorithm~\ref{distributedpolicygradientAlgorithm} becomes slower as $\kappa_p$ increases, which can be attributed to the expanded local policy dependencies and the associated increase in estimation complexity.}
\par
\textbf{Steady-state performance}: As shown in Fig.~\ref{fig:ablation}, the steady-state performance of Algorithm~\ref{distributedpolicygradientAlgorithm} improves monotonically with increasing $\kappa_{p}$.
Notably, the performance improvement from $\kappa_{p} = 0$ to $\kappa_{p}=1$ is substantial, whereas the gain from $\kappa_{p}=1$ to $\kappa_{p}=2$ is relatively marginal.
This is because, under the given simulation setup, one-hop coupling already capture the primary aspects of coordination among agents.

\section{Conclusions}\label{SectionVConclusions}
In this paper, we proposed a novel DSCP algorithm for {\color{blue}the} NMARL problem with coupled policies and proved
{\color{blue}that the gradient norm of the objective function along the joint policy sequence generated by the DSCP algorithm converges to zero both almost surely and in expectation.}
The foundational framework of the DSCP algorithm incorporates the neighbors' averaged $Q$-function and the distributed optimization of policy parameters.
In future work, the use of function approximation techniques for neighbors' averaged $Q$-function in coupled-policy settings will be explored to further improve learning efficiency.

\clearpage
\section{Appendix}\label{SectionAppendix}

\subsection{Proof of Lemma~\ref{thelemmaofdecomposition}}\label{ProofofLemmathelemmaofdecomposition}
\begin{proof}
For the NMARL problem introduced in Section~\ref{ModelofNMARLproblem}, we have a fact that the reward $r_{i,t}=r_{i}(s_{\mathcal{N}_{i},t},a_{\mathcal{N}_{i},t})$ received by agent $i$ depends solely on the state-action pairs of its direct neighbors.
To illustrate this point, for any trajectory of agent $i$ up to an arbitrary time $t'\geq1$ that is generated by the joint policy $\bm{\pi_{\theta}}$, we define $\mathrm{Pr}^{\bm{\pi_{\theta}}}(s_{\mathcal{N}_{i},t'},a_{\mathcal{N}_{i},t'}|\bm{s}_{0},\bm{a}_{0})$ as the probability of the state-action pair $(s_{\mathcal{N}_{i},t'},a_{\mathcal{N}_{i},t'})$ at time $t'$ and  have
\begin{align}
&\mathrm{Pr}^{\bm{\pi_{\theta}}}(s_{\mathcal{N}_{i},t'},a_{\mathcal{N}_{i},t'}|\bm{s}_{0},\bm{a}_{0})\notag\\
=&\sum_{s_{\mathcal{N}_i,1:t'-1},a_{\mathcal{N}_i,1:t'-1}}\prod_{t=1}^{t'}\prod_{j\in\mathcal{N}_{i}}\pi_{j}(a_{j,t}|s_{j,t},\theta_{j},\theta_{\mathcal{N}^{\kappa_{p}}_{j,-j}})\notag\\
&\times\mathcal{P}_{j}(s_{j,t}|s_{j,t-1},a_{j,t-1})\notag\\
=&\mathrm{Pr}^{\bm{\pi_{\theta}}}(s_{\mathcal{N}_{i},t'},a_{\mathcal{N}_{i},t'}|s_{\mathcal{N}_{i},0},a_{\mathcal{N}_{i},0}),\label{addedinlemma1}
\end{align}
where the first equality holds since the global state transition and the joint policy factorize as $\bm{\mathcal{P}}(\bm{s}'|\bm{s},\bm{a})=\prod_{k=1}^{N}\mathcal{P}_k(s'_k|s_k,a_k)$ and $\bm{\pi}_{\bm{\theta}}(\bm{a}|\bm{s})=\prod_{k=1}^{N}\pi_k(a_k|s_k,\theta_k,\theta_{\mathcal{N}^{\kappa_p}_{k,-k}})$, so that summing over the states and actions of agents outside $\mathcal{N}_i$ at times $t=1,\cdots,t'$ yields one. This implies that $(s_{\mathcal{N}_i,t'},a_{\mathcal{N}_i,t'})$ only depends on $(s_{\mathcal{N}_i,0},a_{\mathcal{N}_i,0})$.
Based on this fact, by revisiting the definitions of $Q^{\bm{\pi_{\theta}}}(\bm{s},\bm{a})$ in (\ref{thedefinitionofglobalQfunction}) and $Q^{\bm{\pi_{\theta}}}_{i}(s_{\mathcal{N}_{i}},a_{\mathcal{N}_{i}})$ in (\ref{thelocalQfunctionofglobalpolicy}), we obtain
\begin{align}
Q^{\bm{\pi_{\theta}}}(\bm{s},\bm{a})
=&\mathbb{E}_{\bm{\pi_{\theta}}}\Big[\frac{1}{N}\sum_{t=0}^{\infty}\sum_{i=1}^{N}\gamma^{t}r_{i,t}|\bm{s}_{0}=\bm{s},\bm{a}_{0}=\bm{a}\Big]\notag\\
=&\frac{1}{N}\sum_{i=1}^{N}\mathbb{E}_{\bm{\pi_{\theta}}}\Big[\sum_{t=0}^{\infty}\gamma^{t}r_{i,t}|s_{\mathcal{N}_{i},0}\!=\!s_{\mathcal{N}_{i}},a_{\mathcal{N}_{i},0}\!=\!a_{\mathcal{N}_{i}}\Big]\label{thelemmaofdecomposition1}\\
=&\frac{1}{N}\sum_{i=1}^{N}Q^{\bm{\pi_{\theta}}}_{i}(s_{\mathcal{N}_{i}},a_{\mathcal{N}_{i}}),\label{thelemmaofdecomposition2}
\end{align}
where the second equality holds by (\ref{addedinlemma1}).
Therefore, the lemma is proved.
\end{proof}

\subsection{Proof of Theorem~\ref{thelemmaofthepolicygradienttheorem}}\label{ProofofTheoremthelemmaofthepolicygradienttheorem}
\begin{proof}
By using the policy gradient theorem in~\cite{Sutton2000}, we have
\begin{align}
\nabla_{\theta_{i}}J(\bm{\theta})=&\frac{1}{1-\gamma}\mathbb{E}_{\bm{s}\sim d^{\bm{\pi_{\theta}}}_{\bm{\rho}},\bm{a}\sim\bm{\pi_{\theta}}}[Q^{\bm{\pi_{\theta}}}(\bm{s},\bm{a})\nabla_{\theta_{i}}\log\bm{\pi_{\theta}}(\bm{a}|\bm{s})]\notag\\
=&\frac{1}{1-\gamma}\mathbb{E}_{\bm{s}\sim d^{\bm{\pi_{\theta}}}_{\bm{\rho}},\bm{a}\sim\bm{\pi_{\theta}}}\Big[Q^{\bm{\pi_{\theta}}}(\bm{s},\bm{a})\notag\\
&\times\nabla_{\theta_{i}}\Big(\sum_{j=1}^{N}\log\pi_{j}(a_{j}|s_{j},\theta_{j},\theta_{\mathcal{N}^{\kappa_{p}}_{j,-j}})\Big)\Big]\notag\\
=&\frac{1}{1-\gamma}\mathbb{E}_{\bm{s}\sim d^{\bm{\pi_{\theta}}}_{\bm{\rho}},\bm{a}\sim\bm{\pi_{\theta}}}\Big[Q^{\bm{\pi_{\theta}}}(\bm{s},\bm{a})\notag\\
&\times\sum_{j\in\mathcal{N}^{\kappa_{p}}_{i}}\nabla_{\theta_{i}}\log\pi_{j}(a_{j}|s_{j},\theta_{j},\theta_{\mathcal{N}^{\kappa_{p}}_{j,-j}})\Big]\label{thepolicygradienttheorem1}\\
=&\frac{1}{1-\gamma}\mathbb{E}_{\bm{s}\sim d^{\bm{\pi_{\theta}}}_{\bm{\rho}},\bm{a}\sim\bm{\pi_{\theta}}}\Big[\frac{1}{N}\Big(\sum_{l\in\mathcal{N}^{\kappa_{p}+1}_{i}}Q^{\bm{\pi_{\theta}}}_{l}(s_{\mathcal{N}_{l}},a_{\mathcal{N}_{l}})\notag\\
&+\sum_{l\in\mathcal{N}^{\kappa_{p}+1}_{-i}}Q^{\bm{\pi_{\theta}}}_{l}(s_{\mathcal{N}_{l}},a_{\mathcal{N}_{l}})\Big)\notag\\
&\times\sum_{j\in\mathcal{N}^{\kappa_{p}}_{i}}\nabla_{\theta_{i}}\log\pi_{j}(a_{j}|s_{j},\theta_{j},\theta_{\mathcal{N}^{\kappa_{p}}_{j,-j}})\Big],\label{thepolicygradienttheorem2}
\end{align}
where the equality (\ref{thepolicygradienttheorem1}) comes from the definition of the coupled policy and the last equality can be obtained by (\ref{theequationofDecompositionofvaluefunctions2}).
\par
For any $l\in\mathcal{N}^{\kappa_{p}+1}_{-i}$, it follows that
\begin{align}
&\mathbb{E}_{\bm{s}\sim d^{\bm{\pi_{\theta}}}_{\bm{\rho}},\bm{a}\sim\bm{\pi_{\theta}}}\Big[Q^{\bm{\pi_{\theta}}}_{l}(s_{\mathcal{N}_{l}},a_{\mathcal{N}_{l}})\notag\\
&\times\sum_{j\in\mathcal{N}^{\kappa_{p}}_{i}}\nabla_{\theta_{i}}\log\pi_{j}(a_{j}|s_{j},\theta_{j},\theta_{\mathcal{N}^{\kappa_{p}}_{j,-j}})\Big]\notag\\
=&\sum_{j\in\mathcal{N}^{\kappa_{p}}_{i}}\mathbb{E}_{\bm{s}\sim d^{\bm{\pi_{\theta}}}_{\bm{\rho}},\bm{a}\sim\bm{\pi_{\theta}}}\Big[Q^{\bm{\pi_{\theta}}}_{l}(s_{\mathcal{N}_{l}},a_{\mathcal{N}_{l}})\notag\\
&\times\nabla_{\theta_{i}}\log\pi_{j}(a_{j}|s_{j},\theta_{j},\theta_{\mathcal{N}^{\kappa_{p}}_{j,-j}})\Big]\notag\\
=&\sum_{j\in\mathcal{N}^{\kappa_{p}}_{i}}\sum_{\bm{s},\bm{a}}d^{\bm{\pi_{\theta}}}_{\bm{\rho}}(\bm{s})\prod_{k=1}^{N}\pi_{k}(a_{k}|s_{k},\theta_{k},\theta_{\mathcal{N}^{\kappa_{p}}_{k,-k}})\notag\\
&\times\Big[Q^{\bm{\pi_{\theta}}}_{l}(s_{\mathcal{N}_{l}},a_{\mathcal{N}_{l}})\frac{\nabla_{\theta_{i}}\pi_{j}(a_{j}|s_{j},\theta_{j},\theta_{\mathcal{N}^{\kappa_{p}}_{j,-j}})}{\pi_{j}(a_{j}|s_{j},\theta_{j},\theta_{\mathcal{N}^{\kappa_{p}}_{j,-j}})}\Big]\notag\\
=&\sum_{j\in\mathcal{N}^{\kappa_{p}}_{i}}\sum_{\bm{s},a_{-j}}d^{\bm{\pi_{\theta}}}_{\bm{\rho}}(\bm{s})\prod_{k\neq j}\pi_{k}(a_{k}|s_{k},\theta_{k},\theta_{\mathcal{N}^{\kappa_{p}}_{k,-k}})\notag\\
&\times\Big[Q^{\bm{\pi_{\theta}}}_{l}(s_{\mathcal{N}_{l}},a_{\mathcal{N}_{l}})\sum_{a_{j}}\nabla_{\theta_{i}}\pi_{j}(a_{j}|s_{j},\theta_{j},\theta_{\mathcal{N}^{\kappa_{p}}_{j,-j}})\Big]\notag\\
=&0,\label{thetruncatederror3-1}
\end{align}
where $a_{-j}$ in third equality represents the joint action of all agents except for agent $j$ and
the last two equalities {\color{blue}follow} from two observations.
The first observation is that for $l\in\mathcal{N}^{\kappa_{p}+1}_{-i}$, the neighborhoods $\mathcal{N}_{l}$ and $\mathcal{N}^{\kappa_{p}}_{i}$ are disjoint.
To see this, suppose there exists some agent $k\in\mathcal{N}_{l}\cap\mathcal{N}^{\kappa_{p}}_{i}$. Then the graph distance from $i$ to $l$ would be at most $\kappa_{p}+1$ (via the path $i\rightarrow\cdots\rightarrow k\rightarrow l$), contradicting $l\in\mathcal{N}^{\kappa_{p}+1}_{-i}$.
Consequently, for $j\in\mathcal{N}^{\kappa_{p}}_{i}$, this disjointness implies that
\begin{align}
&\sum_{a_{j}}Q^{\bm{\pi_{\theta}}}_{l}(s_{\mathcal{N}_{l}},a_{\mathcal{N}_{l}})\nabla_{\theta_{i}}\pi_{j}(a_{j}|s_{j},\theta_{j},\theta_{\mathcal{N}^{\kappa_{p}}_{j,-j}})\notag\\
=&Q^{\bm{\pi_{\theta}}}_{l}(s_{\mathcal{N}_{l}},a_{\mathcal{N}_{l}})\sum_{a_{j}}\nabla_{\theta_{i}}\pi_{j}(a_{j}|s_{j},\theta_{j},\theta_{\mathcal{N}^{\kappa_{p}}_{j,-j}}).\notag
\end{align}
The second observation is that by the normalization of the policy, the term $\sum_{a_{j}}\nabla_{\theta_{i}}\pi_{j}(a_{j}|s_{j},\theta_{j},\theta_{\mathcal{N}^{\kappa_{p}}_{j,-j}})$ satisfies
\begin{align}\notag
&\sum_{a_{j}}\nabla_{\theta_{i}}\pi_{j}(a_{j}|s_{j},\theta_{j},\theta_{\mathcal{N}^{\kappa_{p}}_{j,-j}})\notag\\
=&\nabla_{\theta_{i}}\sum_{a_{j}}\pi_{j}(a_{j}|s_{j},\theta_{j},\theta_{\mathcal{N}^{\kappa_{p}}_{j,-j}})=\nabla_{\theta_{i}}1=0.
\end{align}
Substituting (\ref{thetruncatederror3-1}) into (\ref{thepolicygradienttheorem2}), we can obtain (\ref{thepolicygradienttheorem}).
\end{proof}

\subsection{Proof of Theorem~\ref{thelemmainNNMARLITP}}\label{ProofofTheoremthelemmainNNMARLITP}
\begin{proof}
According to (\ref{thepolicygradienttheorem}), we have that
\begin{align}
&\nabla_{\theta_{i}}J(\bm{\theta})\notag\\
=&\frac{1}{1-\gamma}\mathbb{E}_{\bm{s}\sim d^{\bm{\pi_{\theta}}}_{\bm{\rho}},\bm{a}\sim\bm{\pi_{\theta}}}\Big[\frac{1}{N}\!\!\!\!\sum_{l\in\mathcal{N}^{\kappa_{p}+1}_{i}}Q^{\bm{\pi_{\theta}}}_{l}(s_{\mathcal{N}_{l}},a_{\mathcal{N}_{l}})\notag\\
&\times\sum_{j\in\mathcal{N}^{\kappa_{p}}_{i}}\nabla_{\theta_{i}}\log\pi_{j}(a_{j}|s_{j},\theta_{j},\theta_{\mathcal{N}^{\kappa_{p}}_{j,-j}})\Big]\notag\\
=&\frac{1}{1-\gamma}\mathbb{E}_{\bm{s}\sim d^{\bm{\pi_{\theta}}}_{\bm{\rho}},\bm{a}\sim\bm{\pi_{\theta}}}\Bigg[\frac{1}{N}\Big(\!\!\!\sum_{l\in\mathcal{N}^{\kappa_{p}+1}_{i}}\!\!\!\mathbb{E}_{\bm{\pi_{\theta}}}\Big[\sum_{t=0}^{\infty}\gamma^{t}r_{l}(s_{\mathcal{N}_{l},t},a_{\mathcal{N}_{l},t})\Big|\notag\\
&s_{\mathcal{N}_{l},0}=s_{\mathcal{N}_{l}},a_{\mathcal{N}_{l},0}=a_{\mathcal{N}_{l}}\Big]\Big)\notag\\
&\times\sum_{j\in\mathcal{N}^{\kappa_{p}}_{i}}\nabla_{\theta_{i}}\log\pi_{j}(a_{j}|s_{j},\theta_{j},\theta_{\mathcal{N}^{\kappa_{p}}_{j,-j}})\Bigg]\label{theivequationofpolicygradienttheoreminNNMARL-ITPbefore}\\
=&\frac{1}{1-\gamma}\mathbb{E}_{\bm{s}\sim d^{\bm{\pi_{\theta}}}_{\bm{\rho}},\bm{a}\sim\bm{\pi_{\theta}}}\Bigg[\notag\\
&\mathbb{E}_{\bm{\pi_{\theta}}}\Big[\frac{1}{N}\sum_{t=0}^{\infty}\gamma^{t}\sum_{l\in\mathcal{N}^{\kappa_{p}+1}_{i}}r_{l}(s_{\mathcal{N}_{l},t},a_{\mathcal{N}_{l},t})\Big|\notag\\
&s_{\mathcal{N}^{\kappa_{p}+2}_{i},0}=s_{\mathcal{N}^{\kappa_{p}+2}_{i}},a_{\mathcal{N}^{\kappa_{p}+2}_{i},0}=a_{\mathcal{N}^{\kappa_{p}+2}_{i}}\Big]\notag\\
&\times\sum_{j\in\mathcal{N}^{\kappa_{p}}_{i}}\nabla_{\theta_{i}}\log\pi_{j}(a_{j}|s_{j},\theta_{j},\theta_{\mathcal{N}^{\kappa_{p}}_{j,-j}})\Bigg],\label{theivequationofpolicygradienttheoreminNNMARL-ITP}
\end{align}
where the second equality (\ref{theivequationofpolicygradienttheoreminNNMARL-ITPbefore}) is obtained from the definition of the local $Q$-function in (\ref{thelocalQfunctionofglobalpolicy}).
Substituting the definition of $\widehat{Q^{\bm{\pi_{\theta}}}_{i}}(s_{\mathcal{N}^{\kappa_{p}+2}_{i}},a_{\mathcal{N}^{\kappa_{p}+2}_{i}})$ in (\ref{theaction-averagedQfunctionofagenti}) into (\ref{theivequationofpolicygradienttheoreminNNMARL-ITP}), we can derive (\ref{thepolicygradienttheoreminNNMARL-ITP}) directly.
\end{proof}

\subsection{Proof of Lemma~\ref{thelemmaofapproximatedpolicygradient}}\label{Proofofthelemmaofapproximatedpolicygradient}
\begin{proof}
(i) Recalling the definition of $\hat{\nabla}_{\theta_{i}}\Tilde{J}(\bm{\hat{\theta}}_{t})$ in (\ref{thepolicygradientestimation2}), we have
\begin{align}
\|\hat{\nabla}_{\theta_{i}}\Tilde{J}(\bm{\hat{\theta}}_{t})\|=&\frac{1}{1-\gamma}\|\hat{Q}^{\bm{\hat{\pi}}_{\bm{\hat{\theta}}_{t}}}_{i,T_{1}}\hat{g}^{\bm{\hat{\pi}}_{\bm{\hat{\theta}}_{t}}}_{i,T_{1}}\|\notag\\
\leq&\frac{1}{1-\gamma}\Big|\frac{1}{N}\sum_{\tau=0}^{T_{2}}\gamma^{\tau/2}\sum_{j\in\mathcal{N}^{\kappa_{p}+1}_{i}}r_{j,T_{1}+\tau}\Big|\notag\\
&\times\Big\|\!\!\sum_{j\in\mathcal{N}^{\kappa_{p}}_{i}}\!\!\nabla_{\theta_{i}}\log\pi_{j}(a_{j,T_{1}}|s_{j,T_{1}},\hat{\theta}^{i}_{j,t},\hat{\theta}^{i}_{\mathcal{N}^{\kappa_{p}}_{j,-j},t})\Big\|\notag\\
\leq&\frac{BRM_{\kappa_{p}}M_{\kappa_{p}+1}}{(1-\gamma)(1-\gamma^{1/2})N},\label{theapproximatedpolicygradient2}
\end{align}
where the first inequality can be obtained by the definitions of $\hat{Q}^{\bm{\hat{\pi}}_{\bm{\hat{\theta}}_{t}}}_{i,T_{1}}$ and  $\hat{g}^{\bm{\hat{\pi}}_{\bm{\hat{\theta}}_{t}}}_{i,T_{1}}$ in (\ref{theestimateofwildehatQi}) and (\ref{thegradientestimation}), and the last inequality (\ref{theapproximatedpolicygradient2}) follows from Assumptions~\ref{theassumptionofreward}-\ref{theassumptionofpolicy}.
Hence, the (i) of Lemma~\ref{thelemmaofapproximatedpolicygradient} is proved.
\par
(ii) Let $\mathcal{H}_t$ denote the $\sigma$-algebra generated by $\mathcal{F}_t$, the random horizon $T_1$ and the sample trajectory $(\bm{s},\bm{a})_{0:T_1}$ in the $t$-th iteration.
By the definition of $\hat{\nabla}_{\theta_{i}}\Tilde{J}(\bm{\hat{\theta}}_{t})$ in (\ref{thepolicygradientestimation2}) and the tower property of conditional expectation, we have
\begin{align}
&\mathbb{E}[\hat{\nabla}_{\theta_{i}}\Tilde{J}(\bm{\hat{\theta}}_{t})|\mathcal{F}_{t}]\notag\\
=&\frac{1}{1-\gamma}\mathbb{E}\Big[\mathbb{E}[\hat{Q}^{\bm{\hat{\pi}}_{\bm{\hat{\theta}}_{t}}}_{i,T_{1}}|\mathcal{H}_{t}]\Big(\sum_{j\in\mathcal{N}^{\kappa_{p}}_{i}}\nabla_{\theta_{i}}\log\pi_{j}(a_{j,T_{1}}|s_{j,T_{1}},\hat{\theta}^{i}_{j,t},\notag\\
&\hat{\theta}^{i}_{\mathcal{N}^{\kappa_{p}}_{j,-j},t})\Big)\Big|\mathcal{F}_{t}\Big],\label{theunbiasedestimate1}
\end{align}
where we use the fact that the score functions $\{\nabla_{\theta_{i}}\log\pi_{j}(a_{j,T_{1}}|s_{j,T_{1}},\hat{\theta}^{i}_{j,t}\}_{j\in\mathcal{N}^{\kappa_{p}}_{i}}$ only depends on $(s_{\mathcal{N}^{\kappa_p}_i,T_1},a_{\mathcal{N}^{\kappa_p}_i,T_1})$ and $\hat{\bm{\theta}}_t$, and is thus $\mathcal{H}_t$-measurable.
Since $T_2$ is independent of $T_1$ and the sample trajectory, and the subsequent trajectory $(\bm{s},\bm{a})_{T_1+1:T_1+T_2}$ is generated under the fixed executed joint policy $\bm{\hat{\pi}}_{\bm{\hat{\theta}}_t}$ given $\mathcal{F}_t$, by the Markov property, the decoupling property of the state transition function $\mathcal{P}_i(s'_i|s_i,a_i)$ and Theorem~3.4 in~\cite{ZhangSIAM2020}, we have
\begin{align}
\mathbb{E}\big[\hat{Q}^{\bm{\hat{\pi}}_{\bm{\hat{\theta}}_{t}}}_{i,T_{1}}|\mathcal{H}_{t}\big]
=\widehat{Q^{\bm{\hat{\pi}}_{\bm{\hat{\theta}}_{t}}}_{i}}(s_{\mathcal{N}^{\kappa_{p}+2}_{i},T_{1}},a_{\mathcal{N}^{\kappa_{p}+2}_{i},T_{1}}),\label{theconclusionfromzhang}
\end{align}
where the conditional expectation is taken over $T_2$ and $(\bm{s},\bm{a})_{T_1+1:T_1+T_2}$.
Substituting (\ref{theconclusionfromzhang}) into (\ref{theunbiasedestimate1}), and noting that $T_1$ is independent of the sample trajectory given $\mathcal{F}_t$ with $\mathbb{P}(T_1=t')=(1-\gamma)\gamma^{t'}$, we further have
\begin{align}
&\mathbb{E}[\hat{\nabla}_{\theta_{i}}\Tilde{J}(\bm{\hat{\theta}}_{t})|\mathcal{F}_{t}]\notag\\
=&\frac{1}{1-\gamma}\mathbb{E}\Big[\widehat{Q^{\bm{\hat{\pi}}_{\bm{\hat{\theta}}_{t}}}_{i}}(s_{\mathcal{N}^{\kappa_{p}+2}_{i},T_{1}},a_{\mathcal{N}^{\kappa_{p}+2}_{i},T_{1}})\notag\\
&\times\sum_{j\in\mathcal{N}^{\kappa_{p}}_{i}}\nabla_{\theta_{i}}\log\pi_{j}(a_{j,T_{1}}|s_{j,T_{1}},\hat{\theta}^{i}_{j,t},\hat{\theta}^{i}_{\mathcal{N}^{\kappa_{p}}_{j,-j},t})\Big|\mathcal{F}_{t}\Big]\notag\\
=&\frac{1}{1-\gamma}\sum_{t'=0}^{\infty}\mathbb{P}(t'=T_{1})\mathbb{E}\Big[\widehat{Q^{\bm{\hat{\pi}}_{\bm{\hat{\theta}}_{t}}}_{i}}(s_{\mathcal{N}^{\kappa_{p}+2}_{i},t'},a_{\mathcal{N}^{\kappa_{p}+2}_{i},t'})\notag\\
&\times\sum_{j\in\mathcal{N}^{\kappa_{p}}_{i}}\nabla_{\theta_{i}}\log\pi_{j}(a_{j,t'}|s_{j,t'},\hat{\theta}^{i}_{j,t},\hat{\theta}^{i}_{\mathcal{N}^{\kappa_{p}}_{j,-j},t})\Big|\mathcal{F}_{t}\Big]\notag\\
=&\sum_{t'=0}^{\infty}\gamma^{t'}\mathbb{E}\Big[\widehat{Q^{\bm{\hat{\pi}}_{\bm{\hat{\theta}}_{t}}}_{i}}(s_{\mathcal{N}^{\kappa_{p}+2}_{i},t'},a_{\mathcal{N}^{\kappa_{p}+2}_{i},t'})\notag\\
&\times\sum_{j\in\mathcal{N}^{\kappa_{p}}_{i}}\nabla_{\theta_{i}}\log\pi_{j}(a_{j,t'}|s_{j,t'},\hat{\theta}^{i}_{j,t},\hat{\theta}^{i}_{\mathcal{N}^{\kappa_{p}}_{j,-j},t})\Big|\mathcal{F}_{t}\Big]\label{theunbiasedestimate4}\\
=&\frac{1}{1-\gamma}\mathbb{E}_{\bm{s}\sim d^{\bm{\hat{\pi}}_{\bm{\hat{\theta}}_{t}}}_{\bm{\rho}},\bm{a}\sim\bm{\hat{\pi}}_{\bm{\hat{\theta}}_{t}}}\Big[\widehat{Q^{\bm{\hat{\pi}}_{\bm{\hat{\theta}}_{t}}}_{i}}(s_{\mathcal{N}^{\kappa_{p}+2}_{i}},a_{\mathcal{N}^{\kappa_{p}+2}_{i}})\notag\\
&\times\sum_{j\in\mathcal{N}^{\kappa_{p}}_{i}}\nabla_{\theta_{i}}\log\pi_{j}(a_{j}|s_{j},\hat{\theta}^{i}_{j,t},\hat{\theta}^{i}_{\mathcal{N}^{\kappa_{p}}_{j,-j},t})\Big]\label{theunbiasedestimate5}\\
=&\nabla_{\theta_{i}}\Tilde{J}(\bm{\hat{\theta}}_{t}),\notag
\end{align}
where the equality (\ref{theunbiasedestimate4}) is obtained by the fact that $T_{1}$ follows geometric distribution $\mathrm{Geom}(1-\gamma)$ and (\ref{theunbiasedestimate5}) comes from the definition of the discounted state visitation distribution in (\ref{Thediscountedstatevisitationdistribution}).
Hence, the proof of (ii) of Lemma~\ref{thelemmaofapproximatedpolicygradient} is completed.
\end{proof}

\subsection{Proof of Lemma~\ref{thesmoothofpolicygradient}}\label{Proofofthesmoothofpolicygradient}
\begin{proof}
By the definitions of $\nabla_{\theta_{i}}J(\bm{\theta})$ in (\ref{thepolicygradienttheoreminNNMARL-ITP}), we have
\begin{align}
&\nabla_{\theta_{i}}J(\bm{\theta})\notag\\
=&\frac{1}{1-\gamma}\mathbb{E}_{\bm{s}\sim d^{\bm{\pi_{\theta}}}_{\bm{\rho}},\bm{a}\sim\bm{\pi_{\theta}}}\Big[\widehat{Q^{\bm{\pi_{\theta}}}_{i}}(s_{\mathcal{N}^{\kappa_{p}+2}_{i}},a_{\mathcal{N}^{\kappa_{p}+2}_{i}})\notag\\
&\times\sum_{j\in\mathcal{N}^{\kappa_{p}}_{i}}\nabla_{\theta_{i}}\log\pi_{j}(a_{j}|s_{j},\theta_{j},\theta_{\mathcal{N}^{\kappa_{p}}_{j,-j}})\Big]\notag\\
=&\sum_{j\in\mathcal{N}^{\kappa_{p}}_{i}}\sum_{t=0}^{\infty}\sum_{\tau=0}^{\infty}\gamma^{t+\tau}\int\Big(\frac{1}{N}\sum_{l\in\mathcal{N}^{\kappa_{p}+1}_{i}}r_{l,t+\tau}\Big)\notag\\
&\times\nabla_{\theta_{i}}\log\pi_{j}(a_{j,t}|s_{j,t},\theta_{j},\theta_{\mathcal{N}^{\kappa_{p}}_{j,-j}})\rho^{\bm{\pi_{\theta}}}_{t+\tau}(\bm{s}_{0:t+\tau},\bm{a}_{0:t+\tau})\notag\\
&d_{\bm{s}_{0:t+\tau}}d_{\bm{a}_{0:t+\tau}},\label{thepolicygradientsumformal}
\end{align}
where $\rho^{\bm{\pi_{\theta}}}_{t+\tau}(\bm{s}_{0:t+\tau},\bm{a}_{0:t+\tau})$ is the probability of generating sequence $\{\bm{s}_{0:t+\tau},\bm{a}_{0:t+\tau}\}$ under joint policy $\bm{\pi_{\theta}}$ and represented as
\begin{align}\label{theprobabilityofsequence}
&\rho^{\bm{\pi_{\theta}}}_{t+\tau}(\bm{s}_{0:t+\tau},\bm{a}_{0:t+\tau})\notag\\
=&\prod_{h=0}^{t+\tau-1}\bm{\mathcal{P}}(\bm{s}_{h+1}|\bm{s}_{h},\bm{a}_{h})\prod_{h=0}^{t+\tau}\bm{\pi_{\theta}}(\bm{a}_{h}|\bm{s}_{h})\bm{\rho}(\bm{s}_{0}).
\end{align}
By using the representation of $\nabla_{\theta_{i}}J(\bm{\theta})$ in
(\ref{thepolicygradientsumformal}), we have
\begin{align}
&\|\nabla_{\theta_{i}}J(\bm{\theta})-\nabla_{\theta_{i}}J(\bm{\theta}')\|\notag\\
=&\Bigg\|\sum_{j\in\mathcal{N}^{\kappa_{p}}_{i}}\sum_{t=0}^{\infty}\sum_{\tau=0}^{\infty}\gamma^{t+\tau}\Bigg(\int\Big(\frac{1}{N}\sum_{l\in\mathcal{N}^{\kappa_{p}+1}_{i}}r_{l,t+\tau}\Big)\notag\\
&\times\Big(\nabla_{\theta_{i}}\log\pi_{j}(a_{j,t}|s_{j,t},\theta_{j},\theta_{\mathcal{N}^{\kappa_{p}}_{j,-j}})\!\!-\!\!\nabla_{\theta_{i}}\log\pi_{j}(a_{j,t}|s_{j,t},\theta'_{j},\notag\\
&\theta'_{\mathcal{N}^{\kappa_{p}}_{j,-j}})\Big)\rho^{\bm{\pi_{\theta}}}_{t+\tau}(\bm{s}_{0:t+\tau},\bm{a}_{0:t+\tau})d_{\bm{s}_{0:t+\tau}}d_{\bm{a}_{0:t+\tau}}\Bigg)\notag\\
&+\gamma^{t+\tau}\Bigg(\int\Big(\frac{1}{N}\sum_{l\in\mathcal{N}^{\kappa_{p}+1}_{i}}r_{l,t+\tau}\Big)\notag\\
&\times\nabla_{\theta_{i}}\log\pi_{j}(a_{j,t}|s_{j,t},\theta'_{j},\theta'_{\mathcal{N}^{\kappa_{p}}_{j,-j}})\big(\rho^{\bm{\pi_{\theta}}}_{t+\tau}(\bm{s}_{0:t+\tau},\bm{a}_{0:t+\tau})\notag\\
&-\rho^{\bm{\pi}_{\bm{\theta}'}}_{t+\tau}(\bm{s}_{0:t+\tau},\bm{a}_{0:t+\tau})\big)d_{\bm{s}_{0:t+\tau}}d_{\bm{a}_{0:t+\tau}}\Bigg)\Bigg\|.\label{thekeyinsmooth}
\end{align}
Consider that
\begin{align}
&\Big\|\int\Big(\nabla_{\theta_{i}}\log\pi_{j}(a_{j,t}|s_{j,t},\theta_{j},\theta_{\mathcal{N}^{\kappa_{p}}_{j,-j}})\notag\\
&-\nabla_{\theta_{i}}\log\pi_{j}(a_{j,t}|s_{j,t},\theta'_{j},\theta'_{\mathcal{N}^{\kappa_{p}}_{j,-j}})\Big)\rho^{\bm{\pi_{\theta}}}_{t+\tau}(\bm{s}_{0:t+\tau},\bm{a}_{0:t+\tau})\notag\\
&d_{\bm{s}_{0:t+\tau}}d_{\bm{a}_{0:t+\tau}}\Big\|\notag\\
\leq&\int\Big\|\nabla_{\theta_{i}}\log\pi_{j}(a_{j,t}|s_{j,t},\theta_{j},\theta_{\mathcal{N}^{\kappa_{p}}_{j,-j}})\notag\\
&-\nabla_{\theta_{i}}\log\pi_{j}(a_{j,t}|s_{j,t},\theta'_{j},\theta'_{\mathcal{N}^{\kappa_{p}}_{j,-j}})\Big\|\rho^{\bm{\pi_{\theta}}}_{t+\tau}(\bm{s}_{0:t+\tau},\bm{a}_{0:t+\tau})\notag\\
&d_{\bm{s}_{0:t+\tau}}d_{\bm{a}_{0:t+\tau}}\notag\\
\leq&L\|\bm{\theta}-\bm{\theta}'\|,\label{thesmoothpart1-2}
\end{align}
where the last inequality {\color{blue}follows from the Lipschitz continuity of the cross gradient $\nabla_{\theta_i}\log\pi_j$ for $i\in\mathcal{N}_j^{\kappa_p}$} in Assumption~\ref{theassumptionofpolicy}.
By the definition of $\rho^{\bm{\pi_{\theta}}}_{t+\tau}(\bm{s}_{0:t+\tau},\bm{a}_{0:t+\tau})$ in (\ref{theprobabilityofsequence}), we further have
\begin{align}
&|\rho^{\bm{\pi_{\theta}}}_{t+\tau}(\bm{s}_{0:t+\tau},\bm{a}_{0:t+\tau})-\rho^{\bm{\pi}_{\bm{\theta}'}}_{t+\tau}(\bm{s}_{0:t+\tau},\bm{a}_{0:t+\tau})|\notag\\
=&\Big|\bm{\rho}(\bm{s}_{0})\prod_{h=0}^{t+\tau-1}\bm{\mathcal{P}}(\bm{s}_{h+1}|\bm{s}_{h},\bm{a}_{h})\Big(\prod_{h=0}^{t+\tau}\bm{\pi_{\theta}}(\bm{a}_{h}|\bm{s}_{h})\notag\\
&-\prod_{h=0}^{t+\tau}\bm{\pi}_{\bm{\theta}'}(\bm{a}_{h}|\bm{s}_{h})\Big)\Big|\notag\\
\leq&\bm{\rho}(\bm{s}_{0})\prod_{h=0}^{t+\tau-1}\bm{\mathcal{P}}(\bm{s}_{h+1}|\bm{s}_{h},\bm{a}_{h})\Big|\prod_{h=0}^{t+\tau}\bm{\pi_{\theta}}(\bm{a}_{h}|\bm{s}_{h})\notag\\
&-\prod_{h=0}^{t+\tau}\bm{\pi}_{\bm{\theta}'}(\bm{a}_{h}|\bm{s}_{h})\Big|\notag\\
=&\bm{\rho}(\bm{s}_{0})\prod_{h=0}^{t+\tau-1}\bm{\mathcal{P}}(\bm{s}_{h+1}|\bm{s}_{h},\bm{a}_{h})\Big|\int_{0}^{1}(\bm{\theta}-\bm{\theta}')^{\top}\notag\\
&\Big(\sum_{h'=0}^{t+\tau}\nabla_{\bm{\theta}}\bm{\pi}_{\bm{\check{\theta}}_{\xi}}(\bm{a}_{h'}|\bm{s}_{h'})\Big)\prod_{\substack{h=0\\h\neq h'}}^{t+\tau}\bm{\pi}_{\bm{\check{\theta}}_{\xi}}(\bm{a}_{h}|\bm{s}_{h})d_{\xi}\Big|\label{thesmoothpart2-2}\\
\leq&\|\bm{\theta}-\bm{\theta}'\|\int_{0}^{1}\sum_{h'=0}^{t+\tau}\|\nabla_{\bm{\theta}}\log\bm{\pi}_{\bm{\check{\theta}}_{\xi}}(\bm{a}_{h'}|\bm{s}_{h'})\|\notag\\
&\times\rho^{\bm{\pi}_{\bm{\check{\theta}}_{\xi}}}_{t+\tau}(\bm{s}_{0:t+\tau},\bm{a}_{0:t+\tau})d_{\xi},\label{thesmoothpart2-4}
\end{align}
where the equality (\ref{thesmoothpart2-2}) is achieved
by $\bm{\check{\theta}}_{\xi}=\xi\bm{\theta}+(1-\xi)\bm{\theta}'$ with $\xi\in[0,1]$ and the fundamental theorem of calculus.
Consider that
\begin{align}\label{thekeysmoothpart}
&\|\nabla_{\bm{\theta}}\log\bm{\pi}_{\bm{\check{\theta}}_{\xi}}(\bm{a}_{h'}|\bm{s}_{h'})\|\notag\\
=&\begin{Vmatrix}
\nabla_{\theta_{1}}\sum_{j\in\mathcal{N}^{\kappa_{p}}_{1}}\log\pi_{j}(a_{j,h'}|s_{j,h'},\check{\theta}_{\xi,j},\check{\theta}_{\xi,\mathcal{N}^{\kappa_{p}}_{j,-j}})\\
\vdots\\
\nabla_{\theta_{N}}\sum_{j\in\mathcal{N}^{\kappa_{p}}_{N}}\log\pi_{j}(a_{j,h'}|s_{j,h'},\check{\theta}_{\xi,j},\check{\theta}_{\xi,\mathcal{N}^{\kappa_{p}}_{j,-j}})\\
\end{Vmatrix}\notag\\
\leq& BN^{\frac{1}{2}}M_{\kappa_{p}}.
\end{align}
Substituting (\ref{thekeysmoothpart}) into (\ref{thesmoothpart2-4}), we can get
\begin{align}
&|\rho^{\bm{\pi_{\theta}}}_{t+\tau}(\bm{s}_{0:t+\tau},\bm{a}_{0:t+\tau})-\rho^{\bm{\pi}_{\bm{\theta}'}}_{t+\tau}(\bm{s}_{0:t+\tau},\bm{a}_{0:t+\tau})|\notag\\
\leq&
(t+\tau+1)BN^{\frac{1}{2}}M_{\kappa_{p}}\|\bm{\theta}-\bm{\theta}'\|\int_{0}^{1}\rho^{\bm{\pi}_{\bm{\check{\theta}}_{\xi}}}_{t+\tau}(\bm{s}_{0:t+\tau},\bm{a}_{0:t+\tau})d_{\xi}\notag
\end{align}
and
\begin{align}
&\int\big|\rho^{\bm{\pi_{\theta}}}_{t+\tau}(\bm{s}_{0:t+\tau},\bm{a}_{0:t+\tau})\notag\\
&~~~~~~-\rho^{\bm{\pi}_{\bm{\theta}'}}_{t+\tau}(\bm{s}_{0:t+\tau},\bm{a}_{0:t+\tau})\big|d_{\bm{s}_{0:t+\tau}}d_{\bm{a}_{0:t+\tau}}\notag\\
\leq&(t+\tau+1)BN^{\frac{1}{2}}M_{\kappa_{p}}\|\bm{\theta}-\bm{\theta}'\|,\label{thekeysmoothpart1}
\end{align}
which follows from Tonelli's theorem, since $\rho^{\bm{\pi}_{\bm{\check{\theta}}_\xi}}_{t+\tau}$ is a probability distribution for each $\xi\in[0,1]$
\par
Substituting (\ref{thesmoothpart1-2}) and (\ref{thekeysmoothpart1}) into (\ref{thekeyinsmooth}), we further have
\begin{align}
&\|\nabla_{\theta_{i}}J(\bm{\theta})-\nabla_{\theta_{i}}J(\bm{\theta}')\|\notag\\
\leq&\sum_{j\in\mathcal{N}^{\kappa_{p}}_{i}}\sum_{t=0}^{\infty}\sum_{\tau=0}^{\infty}\gamma^{t+\tau}\Big(\frac{LRM_{\kappa_{p}+1}}{N}\|\bm{\theta}-\bm{\theta}'\|\notag\\
&+(t+\tau+1)\frac{B^{2}RM_{\kappa_{p}}M_{\kappa_{p}+1}}{N^{\frac{1}{2}}}\|\bm{\theta}-\bm{\theta}'\|\Big)\notag\\
\leq&\Big(\frac{LRM_{\kappa_{p}}M_{\kappa_{p}+1}}{(1-\gamma)^{2}N}+\frac{(1+\gamma)B^{2}RM^{2}_{\kappa_{p}}M_{\kappa_{p}+1}}{(1-\gamma)^{3}N^{\frac{1}{2}}}\Big)\|\bm{\theta}-\bm{\theta}'\|,\notag
\end{align}
which proves the lemma.
\end{proof}

\subsection{Proof of Corollary~\ref{theerrorbetweenpolicygradient}}
{\color{blue}\begin{proof}
For any $\bm{\hat{\theta}}=((\bm{\hat{\theta}}^1)^\top,\cdots,(\bm{\hat{\theta}}^N)^\top)^\top$, the executed joint policy is $\bm{\hat{\pi}}_{\bm{\hat{\theta}}}(\bm{a}|\bm{s})=\prod_{i=1}^{N}\pi_i(a_i|s_i,\hat{\theta}^{i}_{i},\hat{\theta}^{i}_{\mathcal{N}^{\kappa_p}_{i,-i}})$.
According to (\ref{thepolicygradientestimation1}), we have
\begin{align}
\nabla_{\theta_i}\tilde{J}(\hat{\bm{\theta}})=&\sum_{j\in\mathcal{N}^{\kappa_p}_i}\sum_{t=0}^{\infty}\sum_{\tau=0}^{\infty}\gamma^{t+\tau}\int\Big(\frac{1}{N}\sum_{l\in\mathcal{N}^{\kappa_p+1}_i}r_{l,t+\tau}\Big)\notag\\
&\times\nabla_{\theta_i}\log\pi_j(a_{j,t}|s_{j,t},\hat{\theta}^{i}_{j},\hat{\theta}^{i}_{\mathcal{N}^{\kappa_p}_{j,-j}})\notag\\
&\times\rho^{\bm{\hat{\pi}}_{\bm{\hat{\theta}}}}_{t+\tau}(\bm{s}_{0:t+\tau},\bm{a}_{0:t+\tau})d\bm{s}_{0:t+\tau}d\bm{a}_{0:t+\tau}.\label{corollary1-1}
\end{align}
\par
When $\hat{\bm{\theta}}=\mathbf{1}_N\otimes\bm{\theta}$, we have $\hat{\theta}^{i}_{j}=\theta_j$ for all $i,j\in\mathcal{N}$.
Then $\bm{\hat{\pi}}_{\bm{\hat{\theta}}}=\bm{\pi_{\theta}}$ and the score functions in (\ref{corollary1-1}) coincide with those in (\ref{thepolicygradientsumformal}), which yields $\nabla_{\theta_i}J(\bm{\theta})=\nabla_{\theta_i}\tilde{J}(\mathbf{1}_N\otimes\bm{\theta})$.
\par
For general $\hat{\bm{\theta}}$, following the decomposition in (\ref{thekeyinsmooth}), we have
\begin{align}
&\|\nabla_{\theta_{i}}J(\bm{\theta})-\nabla_{\theta_i}\tilde{J}(\hat{\bm{\theta}})\|\notag\\
=&\Bigg\|\sum_{j\in\mathcal{N}^{\kappa_{p}}_{i}}\sum_{t=0}^{\infty}\sum_{\tau=0}^{\infty}\gamma^{t+\tau}\Bigg(\int\Big(\frac{1}{N}\sum_{l\in\mathcal{N}^{\kappa_{p}+1}_{i}}r_{l,t+\tau}\Big)\notag\\
&\times\Big(\nabla_{\theta_{i}}\log\pi_{j}(a_{j,t}|s_{j,t},\theta_{j},\theta_{\mathcal{N}^{\kappa_{p}}_{j,-j}})\!\!-\!\!\nabla_{\theta_{i}}\log\pi_{j}(a_{j,t}|s_{j,t},\hat{\theta}^{i}_{j},\notag\\
&\hat{\theta}^{i}_{\mathcal{N}^{\kappa_p}_{j,-j}},)\Big)\rho^{\bm{\pi_{\theta}}}_{t+\tau}(\bm{s}_{0:t+\tau},\bm{a}_{0:t+\tau})d_{\bm{s}_{0:t+\tau}}d_{\bm{a}_{0:t+\tau}}\Bigg)\notag\\
&+\gamma^{t+\tau}\Bigg(\int\Big(\frac{1}{N}\sum_{l\in\mathcal{N}^{\kappa_{p}+1}_{i}}r_{l,t+\tau}\Big)\notag\\
&\times\nabla_{\theta_{i}}\log\pi_{j}(a_{j,t}|s_{j,t},\hat{\theta}^{i}_{j},\hat{\theta}^{i}_{\mathcal{N}^{\kappa_p}_{j,-j}})\big(\rho^{\bm{\pi_{\theta}}}_{t+\tau}(\bm{s}_{0:t+\tau},\bm{a}_{0:t+\tau})\notag\\
&-\rho^{\bm{\hat{\pi}}_{\bm{\hat{\theta}}}}_{t+\tau}(\bm{s}_{0:t+\tau},\bm{a}_{0:t+\tau})\big)d_{\bm{s}_{0:t+\tau}}d_{\bm{a}_{0:t+\tau}}\Bigg)\Bigg\|.\label{corollary1-2}
\end{align}
\par
Similar to (\ref{thesmoothpart1-2}), we have
\begin{align}
&\Big\|\int\Big(\nabla_{\theta_{i}}\log\pi_{j}(a_{j,t}|s_{j,t},\theta_{j},\theta_{\mathcal{N}^{\kappa_{p}}_{j,-j}})\notag\\
&-\nabla_{\theta_{i}}\log\pi_{j}(a_{j,t}|s_{j,t},\hat{\theta}^{i}_{j},\hat{\theta}^{i}_{\mathcal{N}^{\kappa_p}_{j,-j}})\Big)\rho^{\bm{\pi_{\theta}}}_{t+\tau}(\bm{s}_{0:t+\tau},\bm{a}_{0:t+\tau})\notag\\
&d_{\bm{s}_{0:t+\tau}}d_{\bm{a}_{0:t+\tau}}\Big\|\notag\\
\leq&L\|\bm{\theta}-\bm{\hat{\theta}}^{i}\|\notag\\
\leq&L\|\mathbf{1}_{N}\otimes\bm{\theta}-\bm{\hat{\theta}}\|.\label{corollary1-3}
\end{align}
\par
Moreover, similar to (\ref{thesmoothpart2-4}) and (\ref{thekeysmoothpart1}), we obtain
\begin{align}
&\int|\rho^{\bm{\pi_{\theta}}}_{t+\tau}(\bm{s}_{0:t+\tau},\bm{a}_{0:t+\tau})\notag\\
&~~~~~~-\rho^{\bm{\hat{\pi}}_{\bm{\hat{\theta}}}}_{t+\tau}(\bm{s}_{0:t+\tau},\bm{a}_{0:t+\tau})|d_{\bm{s}_{0:t+\tau}}d_{\bm{a}_{0:t+\tau}}\notag\\
=&\int|\rho^{\bm{\hat{\pi}}_{\mathbf{1}_{N}\otimes\bm{\theta}}}_{t+\tau}(\bm{s}_{0:t+\tau},\bm{a}_{0:t+\tau})\notag\\
&~~~~~~-\rho^{\bm{\hat{\pi}}_{\bm{\hat{\theta}}}}_{t+\tau}(\bm{s}_{0:t+\tau},\bm{a}_{0:t+\tau})|d_{\bm{s}_{0:t+\tau}}d_{\bm{a}_{0:t+\tau}}\notag\\\notag\\
\leq&\|\mathbf{1}_{N}\otimes\bm{\theta}-\bm{\hat{\theta}}\|\int_{0}^{1}\sum_{h'=0}^{t+\tau}\|\nabla_{\bm{\hat{\theta}}}\log\bm{\hat{\pi}}_{\bm{\acute{\theta}}_{\xi}}(\bm{a}_{h'}|\bm{s}_{h'})\|\notag\\
&\times\rho^{\bm{\hat{\pi}}_{\bm{\acute{\theta}}_{\xi}}}_{t+\tau}(\bm{s}_{0:t+\tau},\bm{a}_{0:t+\tau})d_{\xi}\label{corollary1-4}\\
\leq&(t+\tau+1)BN^{\frac{1}{2}}M_{\kappa_{p}}\|\mathbf{1}_{N}\otimes\bm{\theta}-\bm{\hat{\theta}}\|,\label{corollary1-5}
\end{align}
where (\ref{corollary1-4}) comes from $\bm{\acute{\theta}}_{\xi}=\xi(\mathbf{1}_{N}\otimes\bm{\theta})+(1-\xi)\bm{\hat{\theta}}$ with $\xi\in[0,1]$ and
the last inequality can be obtained by Assumption~\ref{theassumptionofpolicy} and
\begin{align}
&\|\nabla_{\bm{\hat{\theta}}}\log\bm{\hat{\pi}}_{\bm{\acute{\theta}}_{\xi}}(\bm{a}_{h'}|\bm{s}_{h'})\|\notag\\
=&\begin{Vmatrix}
\nabla_{\hat{\theta}^{1}_{1}}\log\pi_{1}(a_{1,h'}|s_{1,h'},\acute{\theta}^{1}_{\xi,1},\acute{\theta}^{1}_{\xi,\mathcal{N}^{\kappa_{p}}_{1,-1}})\\
\vdots\\
\nabla_{\hat{\theta}^{1}_{N}}\log\pi_{1}(a_{1,h'}|s_{1,h'},\acute{\theta}^{1}_{\xi,1},\acute{\theta}^{1}_{\xi,\mathcal{N}^{\kappa_{p}}_{1,-1}})\\
\vdots\\
\nabla_{\hat{\theta}^{N}_{N}}\log\pi_{N}(a_{N,h'}|s_{N,h'},\acute{\theta}^{N}_{\xi,N},\acute{\theta}^{N}_{\xi,\mathcal{N}^{\kappa_{p}}_{N,-N}})
\end{Vmatrix}\leq BN^{\frac{1}{2}}M_{\kappa_{p}}.\notag
\end{align}
\par
Substituting (\ref{corollary1-3}) and (\ref{corollary1-5}) into (\ref{corollary1-2}) and following the same steps as in the proof of Lemma~\ref{thesmoothofpolicygradient}, we obtain (\ref{resultofcorollary1}).
\end{proof}}

\subsection{Proof of Theorem~\ref{thelemmaofpolicyparameterconvergence}}\label{Proofoftheoremthelemmaofpolicyparameterconvergence}
\begin{proof}
For all $i,j\in\mathcal{N}$, denote $\theta_{j,t}(m)$ and $\breve{\theta}^{i}_{j,t}(m)$ be $m$-th parameter of $\theta_{j,t}$ and $m$-th parameter of $\breve{\theta}^{i}_{j,t}$, respectively.
Let $\breve{\theta}_{j,t}(m)=\big(\breve{\theta}^{1}_{j,t}(m),\cdots,\breve{\theta}^{N}_{j,t}(m)\big)^{\top}\in\mathbb{R}^{N}$.
Based on (\ref{thekeyupdateofpolicyparameter-1}), the update of $\breve{\theta}_{j,t}(m)$ can be presented as
\begin{align}\label{thepolicyparameteriterationinm}
\breve{\theta}_{j,t+1}(m)=W\Big(\breve{\theta}_{j,t}(m)+N\big(\theta_{j,t+1}(m)-\theta_{j,t}(m)\big)e_{j}\Big),
\end{align}
where $e_{j}$ is the unit vector with the $j$-th element is 1 and other elements are 0.
Next, we will prove the theorem in two parts: averaged and consensus.
\par
\textbf{Averaged}: Since $W$ is the column stochastic matrix, we multiply both sides of (\ref{thepolicyparameteriterationinm}) by $(1/N)\mathbf{1}^{\top}_{N}$ and have
\begin{align}\label{theaveragedpart}
&\frac{1}{N}\mathbf{1}^{\top}_{N}\breve{\theta}_{j,t+1}(m)\notag\\
=&\frac{1}{N}\mathbf{1}^{\top}_{N}\Big(\breve{\theta}_{j,t}(m)+N\big(\theta_{j,t+1}(m)-\theta_{j,t}(m)\big)e_{j}\Big)\notag\\
=&\frac{1}{N}\mathbf{1}^{\top}_{N}\breve{\theta}_{j,t}(m)+\big(\theta_{j,t+1}(m)-\theta_{j,t}(m)\big).
\end{align}
By adjusting (\ref{theaveragedpart}), we have
\begin{align}\label{theaveragedpart2}
\frac{1}{N}\mathbf{1}^{\top}_{N}\breve{\theta}_{j,t+1}(m)-\theta_{j,t+1}(m)=&\frac{1}{N}\mathbf{1}^{\top}_{N}\breve{\theta}_{j,t}(m)-\theta_{j,t}(m)\notag\\
=&\frac{1}{N}\mathbf{1}^{\top}_{N}\breve{\theta}_{j,1}(m)-\theta_{j,1}(m)\notag\\
=&0,
\end{align}
where the last equality can be obtained by the setting of initial parameters in the Algorithm~\ref{distributedpolicygradientAlgorithm}, i.e., $\breve{\theta}^{i}_{j,1}=\theta_{j,1}=\mathbf{0}_{d}$ for all $i,j\in\mathcal{N}$.
Hence, we have that $(1/N)\mathbf{1}^{\top}_{N}\breve{\theta}_{j,t}(m)=\theta_{j,t}(m)$ for all $t\geq1$.
\par
\textbf{Consensus}: Denote $\delta_{t+1}=WN\big(\theta_{j,t+1}(m)-\theta_{j,t}(m)\big)e_{j}$. Recalling (\ref{thepolicyparameteriterationinm}), the update of $\breve{\theta}_{j,t}(m)$ can be rewritten as
\begin{align}
\breve{\theta}_{j,t+1}(m)=&W\breve{\theta}_{j,t}(m)+\delta_{t+1}\notag\\
=&W^{t}\breve{\theta}_{j,1}(m)+\sum_{k=2}^{t}W^{t-k+1}\delta_{k}+\delta_{t+1}.\label{theconsensuspart1}
\end{align}
Multiplying both sides of formula (\ref{theconsensuspart1}) by matrix $W$ on the left, we obtain
\begin{align}\label{theconsensuspart2}
W\breve{\theta}_{j,t+1}(m)=W^{t+1}\breve{\theta}_{j,1}(m)+\sum_{k=2}^{t+1}W^{t+2-k}\delta_{k}.
\end{align}
Since $W$ is the column stochastic matrix,  we multiply both sides of (\ref{theconsensuspart2}) by $\mathbf{1}^{\top}_{N}$ and have
\begin{align}\label{theconsensuspart3}
\mathbf{1}^{\top}_{N}\breve{\theta}_{j,t+1}(m)=\mathbf{1}^{\top}_{N}\breve{\theta}_{j,1}(m)+\sum_{k=2}^{t+1}\mathbf{1}^{\top}_{N}\delta_{k}.
\end{align}
By multiplying (\ref{theconsensuspart3}) by the vector $\phi_{t+1}$ introduced in Lemma~\ref{thelemmaoftime-varyingnetwork} and subtracting it from (\ref{theconsensuspart2}),
we further have \begin{align}\label{theconsensuspart4}
&(W-\phi_{t+1}\mathbf{1}^{\top}_{N})\breve{\theta}_{j,t+1}(m)\notag\\
=&(W^{t+1}\!-\!\phi_{t+1}\mathbf{1}^{\top}_{N})\breve{\theta}_{j,1}(m)\!+\!\sum_{k=2}^{t+1}(W^{t+2-k}\!-\!\phi_{t+1}\mathbf{1}^{\top}_{N})\delta_{k}.
\end{align}
Denote $\xi_{t+1:k}=W^{t+2-k}-\phi_{t+1}\mathbf{1}^{\top}_{N}$, (\ref{theconsensuspart4}) can be written as
\begin{align}
W\breve{\theta}_{j,t+1}(m)
=&\phi_{t+1}\mathbf{1}^{\top}_{N}\breve{\theta}_{j,t+1}(m)+\xi_{t+1:1}\breve{\theta}_{j,1}(m)\notag\\
&+\sum_{k=2}^{t+1}\xi_{t+1:k}\delta_{k}.\label{theconsensuspart4after}
\end{align}
Recalling the update (\ref{thekeyupdateofpolicyparameter-2}) with $p_{i,1}=1$ for all $i\in\mathcal{N}$, we let $P_{t}=(p_{1,t},\cdots,$ $p_{N,t})^{\top}\in\mathbb{R}^{N}$ and  directly have that $P_{t+1}=WP_{t}=W^{t}P_{1}$ and $\mathbf{1}^{\top}_{N}P_{t+1}=\mathbf{1}^{\top}_{N}P_{t}=\mathbf{1}^{\top}_{N}P_{1}=N$. Furthermore, we have
\begin{align}
P_{t+1}-\phi_{t}\mathbf{1}^{\top}_{N}P_{t+1}=(W^{t}-\mathbf{1}^{\top}_{N}\phi_{t})P_{1}
\end{align}
and
\begin{align}\label{theexpressionofPt}
P_{t+1}=N\phi_{t}+\xi_{t:1}\mathbf{1}_{N}.
\end{align}
Let $\xi_{t:k,i}$ be the $i$-th row element of $\xi_{t:k}$, we can use (\ref{theconsensuspart4after}) and (\ref{theexpressionofPt}) to derive
\begin{align}
&\Big|\hat{\theta}^{i}_{j,t}(m)-\frac{1}{N}\mathbf{1}^{\top}_{N}\breve{\theta}_{j,t}(m)\Big|\notag\\
=&\Bigg|\frac{\phi_{i,t}\mathbf{1}^{\top}_{N}\breve{\theta}_{j,t}(m)+\xi_{t:1,i}\breve{\theta}_{j,1}(m)+\sum_{k=2}^{t}\xi_{t:k,i}\delta_{k}}{N\phi_{i,t}+\xi_{t:1,i}\mathbf{1}_{N}}\notag\\
&-\frac{\mathbf{1}^{\top}_{N}\breve{\theta}_{j,t}(m)}{N}\Bigg|\notag\\
=&\Bigg|\!\frac{N\xi_{t:1,i}\breve{\theta}_{j,1}(m)\!+\!N\sum_{k=2}^{t}\xi_{t:k,i}\delta_{k}\!-\!\xi_{t:1,i}\mathbf{1}_{N}\mathbf{1}^{\top}_{N}\breve{\theta}_{j,t}(m)}{N(N\phi_{i,t}+\xi_{t:1,i}\mathbf{1}_{N})}\!\Bigg|\notag\\
\leq&\frac{\|\xi_{t:1,i}\|\|\breve{\theta}_{j,1}(m)\|+\|\sum_{k=2}^{t}\xi_{t:k,i}\delta_{k}\|}{N\phi_{i,t}+\xi_{t:1,i}\mathbf{1}_{N}}\notag\\
&+\frac{\|\xi_{t:1,i}\mathbf{1}_{N}\|\|\mathbf{1}^{\top}_{N}\breve{\theta}_{j,t}(m)\|}{N(N\phi_{i,t}+\xi_{t:1,i}\mathbf{1}_{N})}\label{theconsensuspart6-1}\\
\leq&\frac{\sqrt{N}\big(M_{1}\lambda^{t-1}\|\breve{\theta}_{j,1}(m)\|+\sum_{k=2}^{t}M_{1}\lambda^{t-k}\|\delta_{k}\|\big)}{\kappa}\notag\\
&+\frac{{\color{blue}\sqrt{N}}M_{1}\lambda^{t-1}\|\breve{\theta}_{j,t}(m)\|}{\kappa}\notag\\
\leq&\frac{\sqrt{N}\sum_{k=2}^{t}M_{1}\lambda^{t-k}\|\delta_{k}\|}{\kappa}+\frac{{\color{blue}\sqrt{N}}M_{1}\lambda^{t-1}\|\breve{\theta}_{j,t}(m)\|}{\kappa}\label{theconsensuspart6-2}
\end{align}
where the first inequality comes from the fact that $|a^{\top}b|\leq\|a\|\|b\|,\forall a,b\in\mathbb{R}^{N}$,
the second inequality can be obtained by Lemma~\ref{thelemmaoftime-varyingnetwork} and the fact that $N\phi_{i,t}+\xi_{t:1,i}\mathbf{1}_{N}\geq\kappa$ (see Corollary~2 in~\cite{NedicTAC2015}), and the last inequality follows from $\breve{\theta}^{i}_{j,1}=\theta_{j,1}=\mathbf{0}_{d}$.
Recalling the setting of $w_{max}>1$ in Section~\ref{subsectionIntermediatelemmas} and (\ref{theconsensuspart1}), we can get
\begin{align}
&\lambda^{t-1}\|\breve{\theta}_{j,t}(m)\|\notag\\
=&\lambda^{t-1}\Big\|W^{t-1}\breve{\theta}_{j,1}(m)+\sum_{k=2}^{t-1}W^{t-k}\delta_{k}+\delta_{t}\Big\|\notag\\
\leq&\lambda^{t-1}\sum_{k=2}^{t-1}w_{max}\|\delta_{k}\|+\|\delta_{t}\|\label{theconsensuspart8-1}\\
\leq&w_{max}\sum_{k=2}^{t}\lambda^{t-k}\|\delta_{k}\|,\label{theconsensuspart8-2}
\end{align}
where the first inequality (\ref{theconsensuspart8-1}) is achieved by the initial $\breve{\theta}^{i}_{j,1}=\mathbf{0}_{d},\forall i,j\in\mathcal{N}$ and the last inequality comes from that $w_{max}>1$ and $\lambda\in(0,1)$.
By the definition of  $\delta_{t}=WN\big(\theta_{j,t}(m)-\theta_{j,t-1}(m)\big)e_{j}$, for all $t\geq2$, we have
\begin{align}
\|\delta_{t}\|=&\|WN\big(\theta_{j,t}(m)-\theta_{j,t-1}(m)\big)e_{j}\|\notag\\
\leq& N\|W\|\|\theta_{j,t}(m)-\theta_{j,t-1}(m)\|\notag\\
\leq& w_{max}N\eta_{\theta,t-1}\|\hat{\nabla}_{\theta_{j}}\Tilde{J}(\bm{\hat{\theta}}_{t-1})\|\notag\\
=&{\color{blue}\frac{w_{max}c_{\theta}N\hat{L}}{t-1}},\label{theconsensuspart7}
\end{align}
where the {\color{blue}last equality} is achieved by {\color{blue}$\eta_{\theta,t}=\frac{c_{\theta}}{t}$} and (i) of Lemma~\ref{thelemmaofapproximatedpolicygradient}.
Substituting (\ref{theconsensuspart8-2}) and (\ref{theconsensuspart7}) into (\ref{theconsensuspart6-2}), we can have
\begin{align}
&\Big|\hat{\theta}^{i}_{j,t}(m)-\frac{1}{N}\mathbf{1}^{\top}_{N}\breve{\theta}_{j,t}(m)\Big|\notag\\
\leq&\frac{{\color{blue}\sqrt{N}(1+w_{max})}M_{1}}{\kappa}\sum_{k=2}^{t}\lambda^{t-k}\|\delta_{k}\|\notag\\
\leq&\frac{{\color{blue}\sqrt{N}(1+w_{max})}w_{max}{\color{blue}c_{\theta}}M_{1}N\hat{L}}{\kappa}\sum_{k=2}^{t}\frac{\lambda^{t-k}}{k-1}\label{theconsensuspart9-1}\\
\leq&\frac{{\color{blue}\sqrt{N}(1+w_{max})}w_{max}{\color{blue}c_{\theta}}M_{1}N\hat{L}}{\kappa}\Big(\frac{1}{t\!-\!1}\sum_{k=2}^{t}\frac{1}{k\!-\!1}\Big)\sum_{k=2}^{t}\lambda^{t-k}\label{theconsensuspart9-2}\\
\leq&\frac{{\color{blue}\sqrt{N}(1+w_{max})}w_{max}{\color{blue}c_{\theta}}M_{1}N\hat{L}}{(1-\lambda)\kappa}\Big(\frac{1}{t-1}\sum_{k=2}^{t}\frac{1}{k-1}\Big)\notag\\
\leq&{\color{blue}\frac{\sqrt{N}(1+w_{max})w_{max}c_{\theta}BRM_{1}M_{\kappa_{p}}M_{\kappa_{p}+1}(1+\log{t})}{(1-\lambda)(1-\gamma)(1-\gamma^{\frac{1}{2}})(t-1)\kappa}},\label{theconsensuspart9-3}
\end{align}
where the inequality (\ref{theconsensuspart9-1}) comes from (\ref{theconsensuspart7}), (\ref{theconsensuspart9-2}) is obtained by the Chebyshev's sum inequality, since $\{\lambda^{t-k}\}$ is increasing and $\{\frac{1}{k-1}\}$ is decreasing in $k$, and the last inequality can be achieved by the setting of $\hat{L}=\frac{BRM_{\kappa_{p}}M_{\kappa_{p}+1}}{(1-\gamma)(1-\gamma^{1/2})N}$ {\color{blue}and the fact that $\sum_{k=2}^{t}\frac{1}{k-1}\leq1+\log{t}$ for all $t\geq2$.}
\par
{\color{blue}Combining the parts of averaged and consensus, i.e., $\theta_{j,t}(m) = (1/N)\mathbf{1}_N^\top \breve{\theta}_{j,t}(m)$ and (\ref{theconsensuspart9-3}), and noting that the policy parameter $\theta_{j,t} \in \mathbb{R}^d$ is $d$-dimensional, we can complete the proof. }
\end{proof}

\subsection{Proof of Theorem~\ref{thetheoremconvergenceofpolicygradient}}\label{Proofoftheoremthetheoremconvergenceofpolicygradient}
\begin{proof}
Define $\hat{\nabla}_{\bm{\theta}}\tilde{J}(\bm{\hat{\theta}}_{t})=\big(\hat{\nabla}_{\theta_{1}}\tilde{J}(\bm{\hat{\theta}}_{t})^{\top},\cdots,$ $\hat{\nabla}_{\theta_{N}}\tilde{J}(\bm{\hat{\theta}}_{t})^{\top}\big)^{\top}$,
by the $L_{2}$-smoothness of $J(\bm{\theta})$ in Lemma~\ref{thesmoothofpolicygradient}, we have
\begin{align}
&J(\bm{\theta}_{t+1})-J(\bm{\theta}_{t})\notag\\
\geq&(\bm{\theta}_{t+1}-\bm{\theta}_{t})^{\top}\nabla_{\bm{\theta}}J(\bm{\theta}_{t})-\frac{1}{2}L_{2}\|\bm{\theta}_{t+1}-\bm{\theta}_{t}\|^{2}\notag\\
\geq&\eta_{\theta,t}\hat{\nabla}_{\bm{\theta}}\tilde{J}(\bm{\hat{\theta}}_{t})^{\top}\nabla_{\bm{\theta}}J(\bm{\theta}_{t})-\frac{1}{2}L_{2}\eta^{2}_{\theta,t}\|\hat{\nabla}_{\bm{\theta}}\tilde{J}(\bm{\hat{\theta}}_{t})\|^{2}\notag\\
\geq&\eta_{\theta,t}\hat{\nabla}_{\bm{\theta}}\tilde{J}(\bm{\hat{\theta}}_{t})^{\top}\nabla_{\bm{\theta}}J(\bm{\theta}_{t})-\frac{1}{2}N\hat{L}^{2}L_{2}\eta^{2}_{\theta,t},\label{theconvergenceof1}
\end{align}
where the second inequality comes from (\ref{theupdateoftruepolicyparameters}) and last inequality can be obtained by the (i) of Lemma~\ref{thelemmaofapproximatedpolicygradient}.
Noting that $\bm{\theta}_t$, $\bm{\hat{\theta}}_t$ and $\nabla_{\bm{\theta}}J(\bm{\theta}_t)$ are $\mathcal{F}_t$-measurable, we have
\begin{align}
&\mathbb{E}[J(\bm{\theta}_{t+1})|\mathcal{F}_{t}]-J(\bm{\theta}_{t})\notag\\
\geq&\eta_{\theta,t}\mathbb{E}[\hat{\nabla}_{\bm{\theta}}\tilde{J}(\bm{\hat{\theta}}_{t})|\mathcal{F}_{t}]^{\top}\nabla_{\bm{\theta}}J(\bm{\theta}_{t})-\frac{1}{2}N\hat{L}^{2}L_{2}\eta^{2}_{\theta,t}\notag\\
=&\eta_{\theta,t}\nabla_{\bm{\theta}}\tilde{J}(\bm{\hat{\theta}}_{t})^{\top}\nabla_{\bm{\theta}}J(\bm{\theta}_{t})-\frac{1}{2}N\hat{L}^{2}L_{2}\eta^{2}_{\theta,t}\label{theconvergenceof2-1}\\
=&\eta_{\theta,t}\big(\nabla_{\bm{\theta}}\tilde{J}(\bm{\hat{\theta}}_{t})-\nabla_{\bm{\theta}}J(\bm{\theta}_{t})+\nabla_{\bm{\theta}}J(\bm{\theta}_{t})\big)^{\top}\nabla_{\bm{\theta}}J(\bm{\theta}_{t})\notag\\
&-\frac{1}{2}N\hat{L}^{2}L_{2}\eta^{2}_{\theta,t}\notag\\
\geq&\eta_{\theta,t}\|\nabla_{\bm{\theta}}J(\bm{\theta}_{t})\|^{2}-\eta_{\theta,t}\big\|\nabla_{\bm{\theta}}\tilde{J}(\bm{\hat{\theta}}_{t})-\nabla_{\bm{\theta}}J(\bm{\theta}_{t})\big\|\notag\\
&\times\big\|\nabla_{\bm{\theta}}J(\bm{\theta}_{t})\big\|-\frac{1}{2}N\hat{L}^{2}L_{2}\eta^{2}_{\theta,t}\notag\\
\geq&\eta_{\theta,t}\|\nabla_{\bm{\theta}}J(\bm{\theta}_{t})\|^{2}-\frac{N^{2}BL_{1}R}{(1-\gamma)^{2}}\eta_{\theta,t}\|\mathbf{1}_{N}\otimes\bm{\theta}_{t}-\bm{\hat{\theta}}_{t}\|\notag\\
&-\frac{1}{2}N\hat{L}^{2}L_{2}\eta^{2}_{\theta,t}\label{theconvergenceof2-2}\\
\geq&\eta_{\theta,t}\|\nabla_{\bm{\theta}}J(\bm{\theta}_{t})\|^{2}-\mathcal{O}(\frac{M^{3}_{\kappa_{p}}M^{2}_{\kappa_{p}+1}\log{t}}{t^{2}})\notag\\
&-\mathcal{O}(\frac{M^{4}_{\kappa_{p}}M^{3}_{\kappa_{p}+1}}{t^{2}}),\label{theconvergenceof2-3}
\end{align}
where the inequality (\ref{theconvergenceof2-1}) comes from (ii) of Lemma~\ref{thelemmaofapproximatedpolicygradient}, the inequality (\ref{theconvergenceof2-2}) can be obtained by Corollary~\ref{theerrorbetweenpolicygradient} and the fact that $\|\nabla_{\bm{\theta}}J(\bm{\theta})\|\leq\frac{N^{\frac{3}{2}}BR}{(1-\gamma)^{2}}$, and the last inequality is achieved by $\hat{L}=\frac{BRM_{\kappa_{p}}M_{\kappa_{p}+1}}{(1-\gamma)(1-\gamma^{1/2})N}$, $L_{1}=\Big(\frac{LRM_{\kappa_{p}}M_{\kappa_{p}+1}}{(1-\gamma)^{2}N}+\frac{(1+\gamma)B^{2}RM^{2}_{\kappa_{p}}M_{\kappa_{p}+1}}{(1-\gamma)^{3}N^{\frac{1}{2}}}\Big)$, $L_{2}=\Big(\frac{LRM_{\kappa_{p}}M_{\kappa_{p}+1}}{(1-\gamma)^{2}N^{\frac{1}{2}}}+\frac{(1+\gamma)B^{2}RM^{2}_{\kappa_{p}}M_{\kappa_{p}+1}}{(1-\gamma)^{3}}\Big)$,
and Theorem~\ref{thelemmaofpolicyparameterconvergence}.
\par
Taking the total expectation on both sides of (\ref{theconvergenceof2-3}) and using the tower property, i.e., $\mathbb{E}\big[\mathbb{E}[J(\bm{\theta}_{t+1})\,|\,\mathcal{F}_t]\big]=\mathbb{E}[J(\bm{\theta}_{t+1})]$, we have
\begin{align}\label{theconvergenceof3}
&\mathbb{E}[J(\bm{\theta}_{t+1})]-\mathbb{E}[J(\bm{\theta}_{t})]\notag\\
\geq&\mathbb{E}[\eta_{\theta,t}\|\nabla_{\bm{\theta}}J(\bm{\theta}_{t})\|^{2}]-\mathcal{O}(\frac{M^{3}_{\kappa_{p}}M^{2}_{\kappa_{p}+1}\log{t}}{t^{2}})\notag\\
&-\mathcal{O}(\frac{M^{4}_{\kappa_{p}}M^{3}_{\kappa_{p}+1}}{t^{2}}).
\end{align}
Summing both sides of (\ref{theconvergenceof3}) for the iterations $t=1,\cdots,T-1$, we have
\begin{align}
&\mathbb{E}[J(\bm{\theta}_{T})]-\mathbb{E}[J(\bm{\theta}_{1})]\notag\\
\geq&\sum_{t=1}^{T-1}\eta_{\theta,t}\mathbb{E}[\|\nabla_{\bm{\theta}}J(\bm{\theta}_{t})\|^{2}]-\sum_{t=1}^{T-1}\mathcal{O}(\frac{M^{3}_{\kappa_{p}}M^{2}_{\kappa_{p}+1}\log{t}}{t^{2}})\notag\\
&-\sum_{t=1}^{T-1}\mathcal{O}(\frac{M^{4}_{\kappa_{p}}M^{3}_{\kappa_{p}+1}}{t^{2}}).\label{theconvergenceof4}
\end{align}
By rearranging (\ref{theconvergenceof4}), it follows that there exists a constant $J_{\mathrm{sup}}(M_{\kappa_{p}},M_{\kappa_{p}+1})$ such that
\begin{align}
&\sum_{t=1}^{T-1}\eta_{\theta,t}\mathbb{E}[\|\nabla_{\bm{\theta}}J(\bm{\theta}_{t})\|^{2}]\notag\\
\leq&\mathbb{E}[J(\bm{\theta}_{T})]-\mathbb{E}[J(\bm{\theta}_{1})]+\sum_{t=1}^{T-1}\mathcal{O}(\frac{M^{3}_{\kappa_{p}}M^{2}_{\kappa_{p}+1}\log{t}}{t^{2}})\notag\\
&+\sum_{t=1}^{T-1}\mathcal{O}(\frac{M^{4}_{\kappa_{p}}M^{3}_{\kappa_{p}+1}}{t^{2}})\notag\\
\leq&J_{\mathrm{sup}}(M_{\kappa_{p}},M_{\kappa_{p}+1}),\label{theconvergenceof5-1}
\end{align}
where the inequality (\ref{theconvergenceof5-1}) comes from the fact that $|J(\bm{\theta})|\leq\frac{R}{1-\gamma}$, and the limits $\lim_{T\rightarrow\infty}\sum_{t=1}^{T-1}\mathcal{O}(\frac{M^{3}_{\kappa_{p}}M^{2}_{\kappa_{p}+1}\log{t}}{t^{2}})$ and $\lim_{T\rightarrow\infty}\sum_{t=1}^{T-1}\mathcal{O}(\frac{M^{4}_{\kappa_{p}}M^{3}_{\kappa_{p}+1}}{t^{2}})$ exist.
\par
{\color{blue}Since $J_{\mathrm{sup}}(M_{\kappa_{p}},M_{\kappa_{p}+1})$ is independent of $T$, letting $T\rightarrow\infty$ in (\ref{theconvergenceof5-1}) yields
\begin{align}
\sum_{t=1}^{\infty}\eta_{\theta,t}\mathbb{E}[\|\nabla_{\bm{\theta}}J(\bm{\theta}_{t})\|^{2}]\leq J_{\mathrm{sup}}(M_{\kappa_{p}},M_{\kappa_{p}+1}).\label{theconvergenceof5-2}
\end{align}
Next, we prove the almost sure convergence and the convergence in expectation of $\|\nabla_{\bm{\theta}}J(\bm{\theta}_{t})\|$ in the following three steps.
\par
\textit{Step 1:} By Tonelli's theorem, the expectation and the summation of nonnegative terms can be interchanged.
(\ref{theconvergenceof5-2}) implies $\mathbb{E}\big[\sum_{t=1}^{\infty}\eta_{\theta,t}\|\nabla_{\bm{\theta}}J(\bm{\theta}_{t})\|^{2}\big]<\infty$, and thus
\begin{align}
\sum_{t=1}^{\infty}\eta_{\theta,t}\|\nabla_{\bm{\theta}}J(\bm{\theta}_{t})\|^{2}<\infty\quad\text{almost surely}.\label{theconvergenceof5-4}
\end{align}
Since (\ref{theconvergenceof5-4}) holds with probability one, it suffices to show that $\lim_{t\rightarrow\infty}\|\nabla_{\bm{\theta}}J(\bm{\theta}_{t})\|=0$ for every realization of the sampled trajectories satisfying (\ref{theconvergenceof5-4}).
To this end, we first note that $\eta_{\theta,t}=\frac{c_{\theta}}{t}$ guarantees $\sum_{t=1}^{\infty}\eta_{\theta,t}=\infty$, which, together with (\ref{theconvergenceof5-4}), implies
\begin{align}
\liminf_{t\rightarrow\infty}\|\nabla_{\bm{\theta}}J(\bm{\theta}_{t})\|=0.\label{theconvergenceof5-4after} \end{align}
Otherwise, there exist $\delta_{0}>0$ and $t_{0}\geq1$ such that $\|\nabla_{\bm{\theta}}J(\bm{\theta}_{t})\|\geq\delta_{0}$ for all $t\geq t_{0}$, which leads to $\sum_{t=t_{0}}^{\infty}\eta_{\theta,t}\|\nabla_{\bm{\theta}}J(\bm{\theta}_{t})\|^{2}\geq\delta^{2}_{0}\sum_{t=t_{0}}^{\infty}\eta_{\theta,t}=\infty$, contradicting (\ref{theconvergenceof5-4}).
\par
\textit{Step 2:}
By (\ref{theupdateoftruepolicyparameters}) and (i) of Lemma~\ref{thelemmaofapproximatedpolicygradient}, for any realization of the sampled trajectories satisfying (\ref{theconvergenceof5-4}), we have
\begin{align}
\|\bm{\theta}_{t+1}-\bm{\theta}_{t}\|=\eta_{\theta,t}\|\hat{\nabla}_{\bm{\theta}}\tilde{J}(\bm{\hat{\theta}}_{t})\|\leq\sqrt{N}\hat{L}\eta_{\theta,t}.\label{theconvergenceof5-5}
\end{align}
Moreover, by the $L_{2}$-smoothness of $J(\bm{\theta})$ in Lemma~\ref{thesmoothofpolicygradient} and the reverse triangle inequality, for any $\bm{\theta},\bm{\theta}'\in\mathbb{R}^{dN}$, we have
\begin{align}
\big|\|\nabla_{\bm{\theta}}J(\bm{\theta})\|-\|\nabla_{\bm{\theta}}J(\bm{\theta}')\|\big|\leq&\|\nabla_{\bm{\theta}}J(\bm{\theta})-\nabla_{\bm{\theta}}J(\bm{\theta}')\|\notag\\
\leq&L_{2}\|\bm{\theta}-\bm{\theta}'\|.\label{theconvergenceof5-gradientnorm}
\end{align}
Suppose, to the contrary, that $\limsup_{t\rightarrow\infty}\|\nabla_{\bm{\theta}}J(\bm{\theta}_{t})\|>\epsilon_{0}$ for some $\epsilon_{0}>0$.
We let $u_{0}=0$ and construct a sequence of intervals $\{[t_{k},u_{k}]\}_{k\geq1}$ recursively.
For $k\geq1$, given $u_{k-1}$, by (\ref{theconvergenceof5-4after}), there exists $t'_{k}>u_{k-1}$ such that $\|\nabla_{\bm{\theta}}J(\bm{\theta}_{t'_{k}})\|<\frac{\epsilon_{0}}{2}$.
Since the limit superior exceeds $\epsilon_{0}$, let $u_{k}$ be the first time after $t'_{k}$ such that $\|\nabla_{\bm{\theta}}J(\bm{\theta}_{u_{k}})\|>\epsilon_{0}$, and let $t_{k}=\max\{t\,|\,t'_{k}\leq t<u_{k},\ \|\nabla_{\bm{\theta}}J(\bm{\theta}_{t})\|<\frac{\epsilon_{0}}{2}\}$.
By construction, $u_{k-1}<t_{k}<u_{k}$, so the intervals are disjoint and $t_{k}\rightarrow\infty$ as $k\rightarrow\infty$. Moreover,
\begin{align}
&\|\nabla_{\bm{\theta}}J(\bm{\theta}_{t_{k}})\|<\frac{\epsilon_{0}}{2},\quad\|\nabla_{\bm{\theta}}J(\bm{\theta}_{u_{k}})\|>\epsilon_{0}\label{theconvergenceof5-intervals}
\end{align}
and
\begin{align}
\|\nabla_{\bm{\theta}}J(\bm{\theta}_{t})\|\geq\frac{\epsilon_{0}}{2},\ \forall t_{k}<t\leq u_{k}.\label{theconvergenceof5-intervals2}
\end{align}
By using (\ref{theconvergenceof5-5}) and (\ref{theconvergenceof5-gradientnorm}), we have
\begin{align}
\frac{\epsilon_{0}}{2}<&\|\nabla_{\bm{\theta}}J(\bm{\theta}_{u_{k}})\|-\|\nabla_{\bm{\theta}}J(\bm{\theta}_{t_{k}})\|\notag\\
\leq& L_{2}\|\bm{\theta}_{u_{k}}-\bm{\theta}_{t_{k}}\|\notag\\
\leq&L_{2}\sum_{t=t_{k}}^{u_{k}-1}\|\bm{\theta}_{t+1}-\bm{\theta}_{t}\|\notag\\
\leq&L_{2}\sqrt{N}\hat{L}\sum_{t=t_{k}}^{u_{k}-1}\eta_{\theta,t},\label{theconvergenceof5-intervals2derive}
\end{align}
which implies $\sum_{t=t_{k}}^{u_{k}-1}\eta_{\theta,t}>\frac{\epsilon_{0}}{2L_{2}\sqrt{N}\hat{L}}$.
Since $\eta_{\theta,t}=\frac{c_{\theta}}{t}$ and $t_{k}\rightarrow\infty$, there exists $v_{0}\geq1$ such that $\eta_{\theta,t_{k}}\leq\frac{\epsilon_{0}}{4L_{2}\sqrt{N}\hat{L}}$ for all $k\geq v_{0}$.
Then, by (\ref{theconvergenceof5-intervals2}), for all $k\geq v_{0}$,
\begin{align}
\sum_{t=t_{k}+1}^{u_{k}}\eta_{\theta,t}\|\nabla_{\bm{\theta}}J(\bm{\theta}_{t})\|^{2}
\geq&\frac{\epsilon^{2}_{0}}{4}\sum_{t=t_{k}+1}^{u_{k}}\eta_{\theta,t}\notag\\
\geq&\frac{\epsilon^{2}_{0}}{4}\sum_{t=t_{k}}^{u_{k}-1}\eta_{\theta,t}-\eta_{\theta,t_{k}}\notag\\
>&\frac{\epsilon^{3}_{0}}{16L_{2}\sqrt{N}\hat{L}}.\label{theconvergenceof5-interval22}
\end{align}
Since the intervals $\{[t_{k}+1,u_{k}]\}_{k\geq v_{0}}$ are disjoint, summing (\ref{theconvergenceof5-interval22}) over all $k\geq v_{0}$, we can derive
\begin{align}
\sum_{t=1}^{\infty}\eta_{\theta,t}\|\nabla_{\bm{\theta}}J(\bm{\theta}_{t})\|^{2}=\infty,\label{theconvergenceof5-interval22after} \end{align}
which contradicts (\ref{theconvergenceof5-4}).
By contradiction, we have
\begin{align}
\limsup_{t\rightarrow\infty}\|\nabla_{\bm{\theta}}J(\bm{\theta}_{t})\|=0.\label{contradictionresult} \end{align}
Combining (\ref{contradictionresult}) with (\ref{theconvergenceof5-4after}),
we have $\lim_{t\rightarrow\infty}\|\nabla_{\bm{\theta}}J(\bm{\theta}_{t})\|=0$ for every realization of the sampled trajectories satisfying (\ref{theconvergenceof5-4}). Since (\ref{theconvergenceof5-4}) holds with probability one, it follows that
\begin{align}
\lim_{t\rightarrow\infty}\|\nabla_{\bm{\theta}}J(\bm{\theta}_{t})\|=0\quad\text{almost surely}.\label{theconvergenceof5-6}
\end{align}

\textit{Step 3:} Recall the definition of $\nabla_{\bm{\theta}}J(\bm{\theta})$ in (\ref{thepolicygradienttheorem}),
we have $\|\nabla_{\bm{\theta}}J(\bm{\theta}_{t})\|\leq\frac{N^{\frac{3}{2}}BR}{(1-\gamma)^{2}}$ for all $t\geq1$, where the upper bound is a finite constant.
By (\ref{theconvergenceof5-6}) and the dominated convergence theorem, we have
\begin{align}\notag
\lim_{t\rightarrow\infty}\mathbb{E}\big[\|\nabla_{\bm{\theta}}J(\bm{\theta}_{t})\|\big]=0.
\end{align}
}
\end{proof}

\vspace*{-1.0cm}

\end{document}